\documentclass[namedreferences,hyperref,optionalrh]{spr-sola}
\usepackage{graphicx}        % For eps figures, newer & more powerfull
\usepackage{amssymb}        % useful mathematical symbols
\usepackage{color}           % For color text: \color command
\usepackage{amsmath}
\usepackage{siunitx}
\usepackage{breakurl}% For breaking URLs easily trough lines in DVI mode
\usepackage{booktabs}
\usepackage{multirow}
\usepackage{tabularx}
\usepackage{enumitem}
\usepackage[rightcaption]{sidecap}
\chardef\us=`\_

\begin{document}

\begin{frontmatter}
\title{Investigating Solar Wind Outflows from Open-Closed Magnetic Field Structures Using Coordinated Solar Orbiter and Hinode Observations}

\author[addressref=aff1,corref,email={nawin.ngampoopun.21@ucl.ac.uk}]{\inits{N.}\fnm{Nawin}~\snm{Ngampoopun}\orcid{0000-0002-1794-1427}}
\author[addressref=aff2,email={roberto.susino@inaf.it}]{\inits{R.}\fnm{Roberto}~\snm{Susino}\orcid{0000-0002-1017-7163}}
\author[addressref={aff3,aff1},email={dhbrooks.works@gmail.com}]{\inits{D.H.}\fnm{David~H.}~\snm{Brooks}\orcid{0000-0002-2189-9313}}
\author[addressref=aff4, email={lionel@predsci.com}]{\inits{R.}\fnm{Roberto}~\snm{Lionello}\orcid{0000-0001-9231-045X}}
\author[addressref=aff2, email={lucia.abbo@inaf.it}]{\inits{L.}\fnm{Lucia}~\snm{Abbo}\orcid{0000-0001-8235-2242}}
\author[addressref=aff5, email={daniele.spadaro@inaf.it}]{\inits{D.}\fnm{Daniele}~\snm{Spadaro}\orcid{0000-0003-3517-8688}}
\author[addressref=aff1, email={deborah.baker@ucl.ac.uk}]{\inits{D.}\fnm{Deborah}~\snm{Baker}\orcid{0000-0002-0665-2355}}
\author[addressref=aff1, email={lucie.green@ucl.ac.uk}]{\inits{L.M.}\fnm{Lucie~M.}~\snm{Green}\orcid{0000-0002-0053-4876}}
\author[addressref={aff6,aff7}, email={david.long@dcu.ie}]{\inits{D.M.}\fnm{David~M.}~\snm{Long}\orcid{0000-0003-3137-0277}}
\author[addressref={aff8,aff1,aff9}, email={steph.yardley@northumbria.ac.uk}]{\inits{S.L.}\fnm{Stephanie~L.}~\snm{Yardley}\orcid{0000-0003-2802-4381}}
\author[addressref=aff1, email={alexander.james@ucl.ac.uk}]{\inits{A.W.}\fnm{Alexander~W.}~\snm{James}\orcid{0000-0001-7927-9291}}
\author[addressref=aff10, email={marco.romoli@unifi.it}]{\inits{M.}\fnm{Marco}~\snm{Romoli}\orcid{0000-0001-9921-1198}}
\author[addressref=aff2, email={silvio.giordano@inaf.it}]{\inits{S.M.}\fnm{Silvio~M.}~\snm{Giordano}\orcid{0000-0002-3468-8566}}
\author[addressref={aff10,aff2}, email={aleksandr.burtovoi@inaf.it}]{\inits{A.}\fnm{Aleksandr}~\snm{Burtovoi}\orcid{0000-0002-8734-808X}}
\author[addressref=aff2, email={federico.landini@inaf.it}]{\inits{F.}\fnm{Federico}~\snm{Landini}\orcid{0000-0001-8244-9749}}
\author[addressref=aff11, email={giuliana.russano@inaf.it}]{\inits{G.}\fnm{Giuliana}~\snm{Russano}\orcid{0000-0002-2433-8706}}
%   NOTE:  Just one corresponding author [corref]
\address[id=aff1]{University College London, Mullard Space Science Laboratory, Holmbury St. Mary, Dorking, Surrey, RH5 6NT, UK}
\address[id=aff2]{National Institute for Astrophysics, Astrophysical Observatory of Torino, Via Osservatorio 20, I-10025 Pino Torinese, Italy}
\address[id=aff3]{Computational Physics Inc., Springfield, VA 22151, USA}
\address[id=aff4]{Predictive Science Inc., San Diego, CA 92121, USA}
\address[id=aff5]{National Institute for Astrophysics, Astrophysical Observatory of Catania, Via Santa Sofia 78, I-95123 Catania, Italy}
\address[id=aff6]{Centre for Astrophysics \& Relativity, School of Physical Sciences, Dublin City University, Glasnevin Campus, Dublin, D09 V209, Ireland}
\address[id=aff7]{Astronomy \& Astrophysics Section, Dublin Institute for Advanced Studies, Dublin D02 XF86, Ireland}
\address[id=aff8]{Department of Mathematics, Physics and Electrical Engineering, Northumbria University, Ellison Place, Newcastle Upon Tyne, NE1 8ST, UK}
\address[id=aff9]{Donostia International Physics Center (DIPC), Paseo Manuel de Lardizabal 4, 20018, San Sebastián, Spain}
\address[id=aff10]{University of Florence, Department of Physics and Astronomy, Via Giovanni Sansone 1, I-50019 Sesto Fiorentino, Italy}
\address[id=aff11]{National Institute for Astrophysics, Astronomical Observatory of Capodimonte, Salita Moiariello 16, I-80131 Napoli, Italy}

\runningauthor{Ngampoopun et al.}
\runningtitle{Investigating Solar Wind Outflows Using Solar Orbiter and Hinode}

\begin{abstract}
ESA/NASA's Solar Orbiter (SO) allows us to study the solar corona at closer distances and from different perspectives, which helps us to gain significant insights into the origins of the solar wind. In this work, we present the analysis of solar wind outflows from two locations: a narrow open-field corridor and a small, mid-latitude coronal hole. These outflows were observed off-limb by the Metis coronagraph onboard SO and on-disk by the Extreme Ultraviolet Imaging Spectrometer (EIS) onboard Hinode. Magnetic field extrapolations suggest that the upflow regions seen in EIS were the sources of the outflowing solar wind observed with Metis. We find that the plasma associated with the narrow open-field corridor has higher electron densities and lower outflow velocities compared to the coronal hole plasma in the middle corona, even though the plasma properties of the two source regions in the low corona are found to be relatively similar. The speed of solar wind from the open-field corridor also shows no correlation with the magnetic field expansion factor, unlike the coronal hole. These pronounced differences at higher altitudes may arise from the dynamic nature of the low-middle corona, in which reconnection can readily occur and may play an important role in driving solar wind variability.
\end{abstract}
\end{frontmatter}
%-------------------------------------------------

% \keywords{Solar Wind; Coronal Holes; Magnetic fields, Corona}
\section{Introduction} \label{sec:intro}

The solar corona continuously expands into interplanetary space and releases streams of plasma and magnetic field in the form of solar wind. Traditionally, the solar wind has been classified by its speed: the fast ($v$ \textgreater\ 450 km s$^{-1}$) and slow ($v$ \textless\ 450 km s$^{-1}$) solar wind. In situ measurements have revealed that these two types of solar wind have distinct properties. The fast solar wind generally has a lower plasma density, higher Alfv{\'e}nicity and lower charge state ratio, implying a lower electron temperature at the source regions than the slower counterpart \citep{Geiss1995}. The fast wind also does not have a significant enhancement in the abundance of elements with a low first-ionisation potential \citep[FIP;][]{Laming2015}, which is in contrast to the enhancement commonly found in the slow solar wind. It is also worth noting that slow solar wind streams have more variability in their properties, with some portions even behaving like a fast solar wind \citep[i.e., slow alfv{\'e}nic solar wind;][]{D'Amicis2015, Stansby2020, Yardley2024}.
%possible source region

Despite the fact that solar wind streams have been studied for several decades, their origin and acceleration mechanisms still remain among the major open questions in heliophysics \citep[e.g.,][]{Viall2020}. The fast solar wind is generally accepted to originate from the core of coronal holes (CHs), regions of the solar corona with relatively low plasma density and temperature, appearing dark in extreme ultraviolet (EUV) and soft X-ray images. %In particular, plasma outflows are observed along open magnetic flux tubes that root in the supergranular network boundaries \citep{Hassler1999, Tu2005} within the CHs. 
The origin of slow solar wind, on the other hand, is still under heavy debate \citep[e.g.,][]{Abbo2016}. Although the slow Alfv{\'e}nic wind is thought to originate from small, low-latitude CH or CH boundaries \citep{Wang2019, D'Amicis2021}, the more variable slow solar wind, characterised by low-FIP element enhancements and higher charge state ratios, is argued to originate from hotter coronal regions, such as at the peripheries of active regions (ARs), where widespread plasma upflows are evident \citep{Sakao2007, Brooks2011, Brooks2015}. Off-limb observations show that the slow solar wind emerges from extended coronal streamers observed in visible light, either from the open field at the streamer edges \citep{Susino2008, Abbo2010}, or as blobs from reconnection at the streamer cusps \citep{Einaudi1999, Sheeley2009}. The variable nature of the solar wind can also arise from rapid changes in magnetic connectivity to multiple source regions with different properties \citep{Yardley2024}.

%Interchange reconnection and S-web model
Interchange reconnection between open and closed magnetic field lines has been suggested to be a mechanism for the release of solar wind plasma \citep{Crooker2002, Fisk2003}. This process allows plasma trapped in closed field regions, such as ARs or coronal streamers, to be released along the (newly formed) open field, which may explain the coronal compositional signature of the slow solar wind \citep{Baker2009, vanDriel-Gesztelyi2012, Brooks2015, Yardley2024}. Interchange reconnection may also be responsible for the release of the fast solar wind, as shown by the observation of small-scale coronal jets in CHs \citep{Chitta2023a, Raouafi2023, Long2023}, and the magnetic field reversals found in the solar wind in the inner heliosphere \citep{Bale2019, Bale2023}.

In particular, \citet{Antiochos2011} proposed that large-scale interchange reconnection takes place at the boundary between closed and open field regions, which can be mapped to the network of separatrix surfaces and quasi-separatrix layers \citep[QSLs;][]{Demoulin1996} called the S-web. The S-web is theoretically defined as a set of arcs with drastic changes in magnetic connectivity found throughout the solar corona. The photospheric footpoints of the S-web structures are located at the boundaries of CHs, the peripheries of ARs, and narrow, sometimes infinitesimal, open-field corridors that link disconnected CHs \citep{Titov2011, Higginson2017, Scott2019}. Structures such as helmet streamers or pseudostreamers are also intrinsically related to the S-web, as they naturally give rise to open-closed field boundaries and magnetic null points \citep{Wang2007, Titov2011, Antiochos2011}. 

Recently, \citet{Chitta2023b} presented observations of complex elongated plasma features in the off-limb EUV and visible light observations. These elongated plasma structures were interpreted as the imprints of the S-web above the CH-AR system and also as observational evidence of the slow solar wind streams. \citet{Baker2023} investigated plasma upflows associated with a thin open field corridor embedded in an AR, which were then associated through magnetic connectivity mapping with variable slow solar wind streams. These results were interpreted as supporting evidence of the framework that reconnection dynamics along the S-web are responsible for the release of (at least part of) slow solar wind plasma. 

%off-limb observation UVCS and middle corona to look at the acceleration of solar wind
The acceleration of the solar wind, as well as the transition from a closed to an open field configuration, occurs in the middle corona, here defined as the region with a heliocentric distance of 1.5 to 6 R$_{\odot}$ \citep{West2023}. Previous coronagraph observations showed significant differences in the electron density and solar wind outflow speeds in equatorial coronal streamers compared to polar CHs in the middle corona, with polar CHs having lower density and higher solar wind speed \citep{Antonucci2004,Antonucci2005,Abbo2010}. The outflow speeds seem to correspond to the degree of super-radial expansion of open magnetic flux tubes \citep{Wang1990}. In addition, preferential heating and acceleration of the ions are observed in both the fast solar wind from CHs \citep{Cranmer2008} and (to a lesser extent) the slow solar wind from streamers \citep{Spadaro2007, Abbo2010}, suggesting the effect of kinetic-scale physics, such as ion cyclotron resonant waves, \citep[see reviews by, e.g.,][]{Antonucci2006, Cranmer2009, Cranmer2019}. 

The complex and dynamic nature of the corona has been revealed by high spatiotemporal off-limb observations. The fine-scale structures of coronal plasma and magnetic field in the low corona can be identified from total solar eclipse observations in visible light \citep{Habbal2011, Habbal2021, Druckmuller2014}, as well as EUV imaging \citep{Seaton2021, Morton2023}. Higher up in the middle and extended corona, similar structures are found to constantly propagate outward from the Sun and impose themselves in the interplanetary solar wind, as shown by coronagraph observations \citep{DeForest2018, Alzate2021}. The density variations and reconfigurations of these fine-scale structures may imply that ongoing reconnection is taking place, which can also contribute to the energisation and acceleration of the solar wind \citep{Chitta2023b, Liewer2023, Ventura2023}.

In this paper, we attempt to provide new insights into the origin of the solar wind using coordinated remote sensing observations between the Solar Orbiter \citep[SO;][]{Muller2020} spacecraft and Earth-orbiting satellites, which are the Solar Dynamics Observatory \citep[SDO;][]{Pesnell2012} and Hinode \citep{Kosugi2007}. We simultaneously derive properties of solar wind plasma emanating from two open-field regions in the low corona (using on-disk spectroscopy) and in the middle corona (using off-limb coronagraph observations), which we link together using magnetic extrapolations. The paper is structured as follows. The instruments and datasets used in this study are described in Section \ref{sec:obs}. The magnetic extrapolation methods and the global magnetic structure of the solar corona are detailed in Section \ref{sec:model}. In Sections \ref{sec:res1} and \ref{sec:res2}, we present the main results from the analysis of the solar wind in the low coronal and the middle coronal observations, respectively. Section \ref{sec:res3} shows the evolution and dynamics of the low and middle corona. Finally, we discuss the results and summarise our findings in Section \ref{sec:disc}.

\section{Remote Sensing Observations} \label{sec:obs}

On 2023 April 9, Solar Orbiter reached perihelion at a heliocentric distance of 0.29 AU and was located around 60$^{\circ}$ west of the Sun-Earth line. This configuration allowed us to observe the solar corona using remote sensing instruments from two different viewpoints: Earth-based satellites (SDO and Hinode) and SO. Note that the difference in distances of the individual spacecraft from the Sun meant that phenomena were observed at different local spacecraft time. Hence, to avoid confusion, we will use the time at Earth for all observations in this paper.

We investigated the solar wind outflows in the middle corona using the Metis coronagraph \citep{Antonucci2020, Fineschi2020} onboard SO. Metis provides coronagraph observations of the off-limb corona in an annular field of view (FOV) ranging between 1.6 -- 2.9$^{\circ}$ (corresponding to 1.76 -- 3.75~R$_{\odot}$ at a heliocentric distance of 0.29~AU) in two passbands: visible light (VL; 580 -- 640~nm) and ultraviolet neutral hydrogen Ly$\alpha$ (UV; $121.6 \pm 10$~nm). From 04:55~UT to 23:55~UT, Metis acquired sequences of VL total brightness (tB) and polarised brightness (pB) images in parallel with UV Ly$\alpha$ images, with a temporal cadence of 20~min, a total effective exposure time of 15~min for both channels, and spatial resolutions of 20\SI{}{\arcsecond} per pixel in the VL and 40\SI{}{\arcsecond} per pixel in the UV. We used level 2 data, which were calibrated using the most up-to-date calibration available and can be accessed through the Solar Orbiter Archive (SOAR)\footnote{\url{https://doi.org/10.5270/esa-366ut35}}. Standard calibration operations include correction for detector bias, dark current, and flat field, optical vignetting function, and radiometric calibration \citep{DeLeo2023,DeLeo2024}.

We also used extreme-ultraviolet (EUV) images in the 174 {\AA} passband from the Full Sun Imager (FSI) telescope of the Extreme Ultraviolet Imager \citep[EUI;][]{Rochus2020} onboard SO, to investigate the low-coronal structure in the region below Metis's inner FOV and complement middle corona observations. This telescope takes full solar disc images with a plate scale of 4.4\SI{}{\arcsecond} per pixel and a time cadence of 10~min. We used calibrated level 2 EUI FITS files from EUI data release 6 \citep{euidatarelease6} for this analysis.

From Earth's viewpoint, we investigated the plasma dynamics in the regions directly below Metis's FOV. As detailed in Hinode Operation Plan 462\footnote{\url{https://www.isas.jaxa.jp/home/solar/hinode_op/hop.php?hop=0462}}, the EUV Imaging Spectrometer \citep[EIS;][]{Culhane2007} onboard the Hinode satellite constructed two consecutive column mosaics approximately 30$^{\circ}$ east of the central meridian, that correspond to the east solar limb as seen by SO. In this observing campaign, two EIS studies were used. The first study was DHB\_007\_v2, a raster scan with a FOV size of 248\SI{}{\arcsecond}$\times$ 512\SI{}{\arcsecond}. The slit size was 2\SI{}{\arcsecond} with an exposure time of 60~s, and rastered in 4\SI{}{\arcsecond} steps. The second study used was CH\_bound\_240x512v1, a raster scan with a FOV size of 240\SI{}{\arcsecond}$\times$ 512\SI{}{\arcsecond}. The slit size was also 2\SI{}{\arcsecond} and rastered in 4\SI{}{\arcsecond} steps, but each slit had an exposure time of 100~s.

Two of the 12 total scans were used for the analysis in this work. The first scan used the DHB\_007\_v2 study and was run from 07:11~UT to 08:15~UT centred at (x, y) $\sim$ (-473\SI{}{\arcsecond}, -607\SI{}{\arcsecond}). The second scan used the CH\_bound\_240x512v1 study centred at (x, y) $\sim$ (-475\SI{}{\arcsecond}, 617\SI{}{\arcsecond}) and was run from 12:13~UT -- 13:55~UT. The obtained spectra were corrected for instrumental effects, including slit tilt, orbital variation, dark current, and warm/hot/dusty pixels before further analysis. Plasma dynamics and properties were then obtained by fitting the spectral data using the EISPAC Python library \citep{Weberg2023}.

Lastly, to provide context for the overall structure and investigate magnetic properties of the solar wind source regions, we used observations from the Atmospheric Imaging Assembly \citep[AIA;][]{Lemen2012} and the Helioseismic and Magnetic Imager \citep[HMI;][]{Scherrer2012} onboard SDO. AIA continuously monitors the full solar disk in seven EUV passbands, with a plate scale of 0.6\SI{}{\arcsecond} per pixel and 12~s cadence. On the other hand, HMI provides line-of-sight (LOS) photospheric magnetograms with a temporal resolution of 45~s and a plate scale resolution of 0.505\SI{}{\arcsecond} per pixel. %The photon noise level is approximately 7 G \citep{Couvidat2016}.
The level 1 AIA observations were processed using a standard routine in the aiapy Python library \citep{Barnes2020}. The HMI magnetograms were then coaligned with the AIA observations.
\begin{figure}[h!]
    \includegraphics[width=\textwidth]{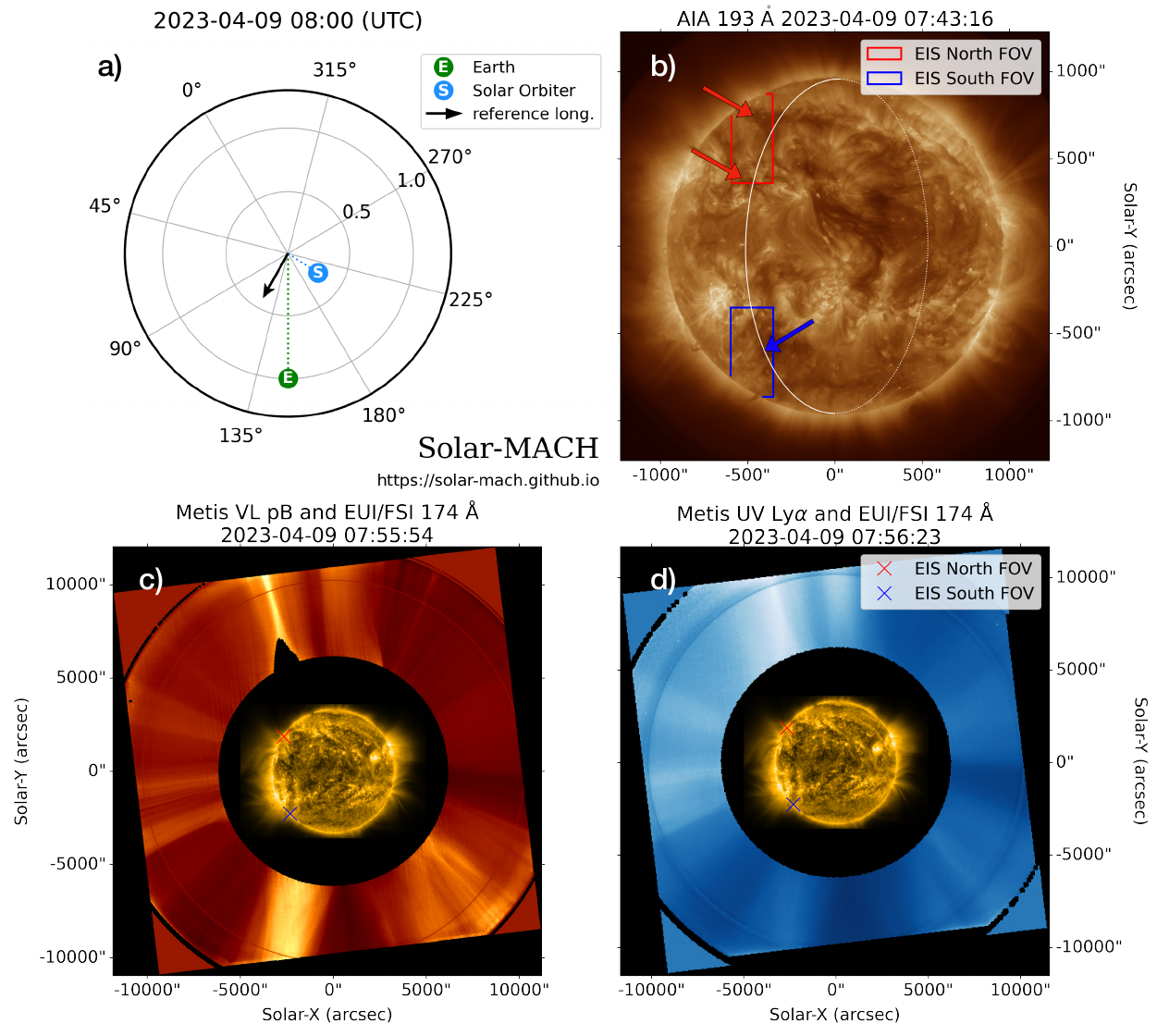}
    \caption{Overview of the remote sensing observations made by SO, Hinode and SDO on 2023 April 9. (a) The diagram showing position of Earth (equivalent to SDO and Hinode) and SO in Carrington coordinate system. The black arrow indicate the Carrington longitude of the solar east limb seen by SO. (b) AIA 193 {\AA} image with the solar limb seen by SO is plotted as a white line. The red arrows point to the solar filaments, and the blue arrow points to a small CH. (c) The composite of EUI/FSI 174 {\AA} and Metis VL pB observations. (d) The composite of EUI/FSI 174 {\AA} and Metis UV Ly$\alpha$ observations. The EIS FOVs are shown in coloured boxes and crosses, red for the North FOV and blue for the South FOV.}
    \label{fig:Fig1_AIAandSO}
\end{figure}

Figure \ref{fig:Fig1_AIAandSO} provides an overview of the remote sensing observations we used for this analysis. Panel a shows the top view of the position of Earth and SO during the observation period generated from the SOLAR-MACH tool \citep{Gieseler2023}, indicating that SO was located at $\sim60^{\circ}$ west of the Sun-Earth line and the east limb (black arrow) seen from SO can be observed on the solar disc by Earth orbiting Hinode and SDO.

Panel b displays the solar corona observed from Earth's perspective using AIA. The FOVs of two EIS raster scans are shown in the red and blue boxes in the left panel, which partly overlap the location of the solar limb as seen by SO, denoted as a white line. The EIS South FOV (blue) corresponds to a small mid-latitude coronal hole (CH; blue arrow), which was identified as a relatively dark region on the solar disc in AIA 193~\AA\ observations. The EIS North FOV (red), on the other hand, observes a relatively bright region lying between two solar filaments, denoted by two red arrows. We interpret that this region corresponds to a decayed AR, as NOAA AR 12331 was observed at the same exact location on 2023 February 14, two solar rotations prior to our observations.

Panels c and d present the observations of the low and middle corona by EUI and Metis onboard SO. The pB and UV observations are enhanced using the Normalising-Radial-Graded Filtering method \citep[NRGF;][]{Morgan2006}. The red and blue crosses correspond to the centre of the EIS North and South FOVs, confirming that they were observing the east solar limb region seen from SO as planned.  The middle corona region corresponding to the EIS South FOV has a lower emission in the VL pB and UV Ly$\alpha$ intensity than that above the EIS North FOV, as observed from Metis.

\section{Global Magnetic Configuration} \label{sec:model}
The structure and plasma dynamics in the low and middle corona are directly related to the global magnetic field configuration. Therefore, we employ a magnetic field extrapolation based on Predictive Science's Magnetohydrodynamics Around a Sphere simulation code \citep[PSI-MAS;][]{Mikic1999} for this analysis. This method solves a set of resistive magnetohydrodynamics (MHD) equations in a 3D non-uniform spherical coordinate grid covering a heliocentric distance of 1 -- 30~R$_{\odot}$ to simulate the volume-filling plasma and magnetic field states of the solar corona. A synoptic photospheric magnetogram is used to provide the boundary condition for the radial magnetic field at 1~R$_{\odot}$. Various known processes in the solar corona, such as radiative losses, anisotropic thermal conduction, and coronal heating, have also been accounted for in the simulation with several degrees of sophistication \citep{Riley2021}. The PSI-MAS model is generally consistent with total solar eclipse observations \citep[e.g.,][]{Boe2021, Boe2023}, which demonstrates that the model can reasonably predict the structure in the low and middle corona. A more detailed description of the PSI-MAS model can be found in \citet{Mikic1999}, \citet{Lionello2009}, and \citet{Mikic2018}.

In our analysis, we used the semi-empirical thermodynamic model of the PSI-MAS simulation described in \citet{Lionello2009}, which accounts for the energy transport processes and plasma dynamics using an empirical heating function. The HMI radial synoptic map from Carrington rotation 2269 (CR2269; 2023 March 24--April 20) is chosen as the lower boundary condition. The simulation results are publicly available from Predictive Science's website\footnote{\url{https://www.predsci.com/mhdweb/data_access.php}}, which includes the three components of the magnetic field and several important plasma properties. We traced magnetic field lines from the photosphere up to 30~R$_{\odot}$, allowing us to investigate magnetic field structures without common constraints such as the source surface height. 

\begin{figure}[t!]
    \centering
    \includegraphics[width = \textwidth]{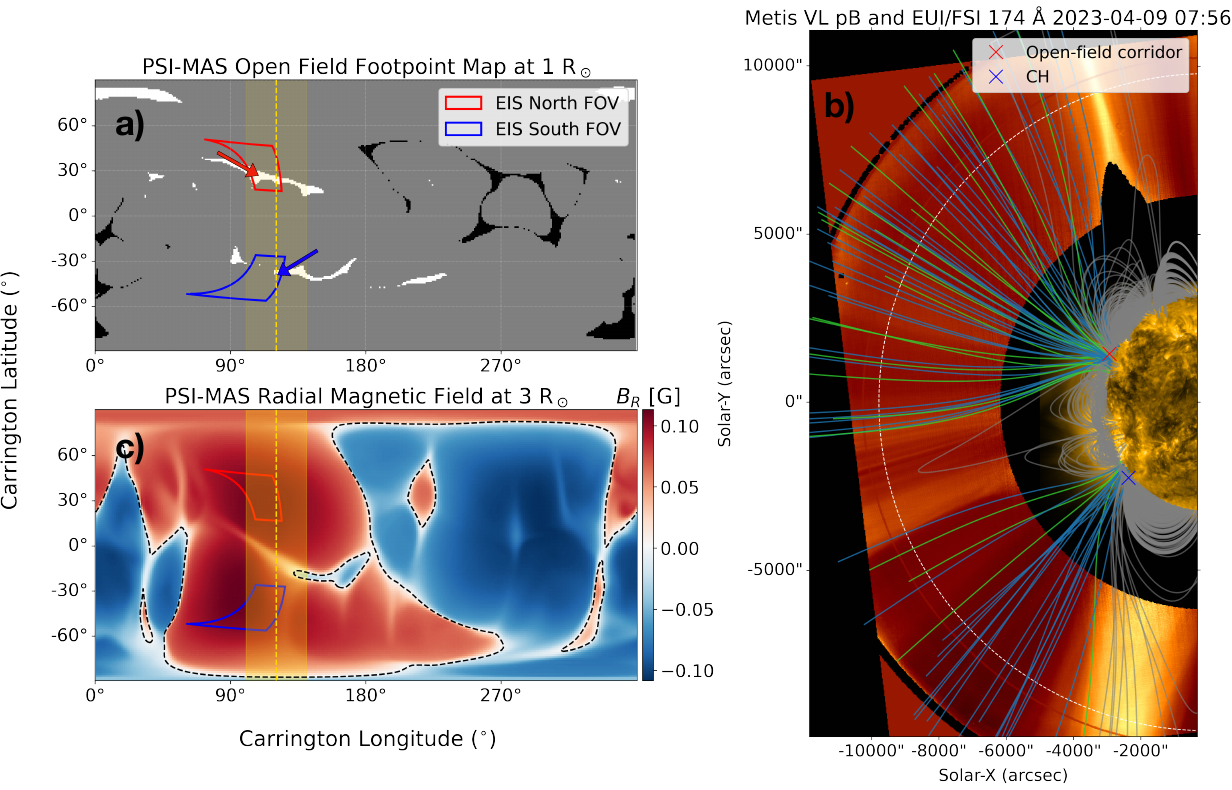}
    \caption{The magnetic field extrapolations derived from the PSI-MAS MHD simulation during Carrington rotation 2269. (a) Carrington synoptic map showing the footpoints of the positive (white) and negative (black) polarity of the open field. The yellow shaded region indicates the area corresponding to the east solar limb seen by Metis, from where the field lines in panel b) were traced from. The red and blue arrows point to the open field footpoints corresponding to a thin open-field corridor and a small CH, respectively. (b) Extrapolated magnetic field lines plotted over the composite EUI/FSI 174~{\AA} and Metis VL pB observations. The grey lines indicate the closed field lines. The light green lines highlight the open field lines with an expansion factor $f_s$ \textgreater\ 20, and the blue field lines shows those with $f_s$ \textless\ 20 (see text). The red and blue crosses denote the location of the open-field corridor and the CH. The dashed white line denotes a heliocentric distance of 3~R$_\odot$. (c) Carrington synoptic map of the radial magnetic field at 3~R$_\odot$. Positive (negative) polarities are shown in red (blue). The polarity inversion lines separating positive and negative polarities are indicated by black dashed lines, with the longest line corresponds to the heliospheric current sheet.}
    \label{fig:Fig2_MAS}
\end{figure}

Figure \ref{fig:Fig2_MAS} summarises the results from the magnetic field extrapolations. Panel a shows the locations of the open field at the photosphere for CR2269. The east limb as seen by Metis and surrounding regions ($\pm$20$^{\circ}$) are denoted as yellow dashed lines and shaded regions. The locations of EIS's FOVs (red and blue boxes) overlap with the footpoints of positive-polarity open magnetic fields (white regions). In particular, the EIS North FOV corresponds to a thin open-field corridor at Carrington latitude $\sim 30 ^{\circ}$, indicated by a red arrow. The EIS South FOV, on the other hand, corresponds to the CH, indicated by a blue arrow (see also Figure \ref{fig:Fig1_AIAandSO}). This strongly suggests that EIS was observing the source regions of the solar wind. Hence, if there are any plasma upflows along these open-field regions observed by EIS, they are likely to be the low-coronal origin of the solar wind.

Panel b of Figure \ref{fig:Fig2_MAS} illustrates the selected extrapolated field lines overlaid on the composite Metis VL pB and EUI/FSI 174~\AA\ observations. All plotted field lines have footpoints within the yellow-shaded region in panels a and c. Near the equator, we can see that there are closed magnetic fields associated with a large pseudostreamer. This structure is surrounded by two open-field regions in the north and south, which are rooted in the open-field footpoints inside the FOVs of the EIS observations, i.e., the open-field corridor (red cross) and the CH (blue cross). The extrapolated magnetic field lines match well with the structure of the middle corona observed by Metis. In particular, the open-field structure in the south corresponds to the dark region seen in Metis pB observations, associated with a CH region. The northern open-field region, on the other hand, appears brighter in pB observations compared to the region corresponding to the southern CH. %The differences in pB intensity from two open-field regions suggested that the corresponding solar wind might have different properties in the middle corona.

Panel c shows the derived radial magnetic field strength in the middle corona at 3~R$_{\odot}$. At this height, the east solar limb seen by SO is mainly filled by positive open magnetic fields that expand from the photosphere. The heliospheric current sheet, defined as the boundary between positive and negative open field regions, is highly inclined (shown by the longest black dashed line), indicating that the solar magnetic field configuration during this time is much more complex compared to the simple dipolar field during solar minimum. The helmet streamer near the south pole seen in panel b also roughly marks the location of the heliospheric current sheet.

We then quantify the properties of the coronal magnetic field using two parameters. The first parameter is the expansion factor, which measures the degree of super-radial expansion of open magnetic flux tubes. The expansion factor $f_s$ is defined as \citep{Wang1990, Antonucci2023},
\begin{equation}
    f_s = \left(\frac{\text{R}_{\odot}}{r_1} \right)^2 \frac{B_R(\text{R}_{\odot})}{B_R(r_1)}
\end{equation}
where $B_R(\text{R}_\odot)$ and $B_R(r_1)$ are the radial magnetic field strengths at the photosphere and at a specific radial distance $r_1$ in the corona. Using the potential field source surface extrapolations and the measured solar wind speed at 1~AU, \citet{Wang1990} found an inverse correlation between the solar wind speed and $f_s$, where $f_s$ is computed at $r_1$ = 2.5~R$_\odot$ (that is, the radius of the source surface). In general, a low expansion factor corresponds to the fast solar wind speeds, whereas high expansion factors correspond to slow solar wind speeds. We illustrate the expansion factors associated with open field lines in panel b of Figure \ref{fig:Fig2_MAS}. The light green lines indicate open field lines with $f_s$ greater than 20 at $r_1$ = 3~R$_\odot$, while the blue lines indicate those with $f_s$ less than 20. This criterion is similar to that defined in \citet{Wang2019}.

\begin{figure}[t!]
    \centering
    \includegraphics[width=0.7\textwidth]{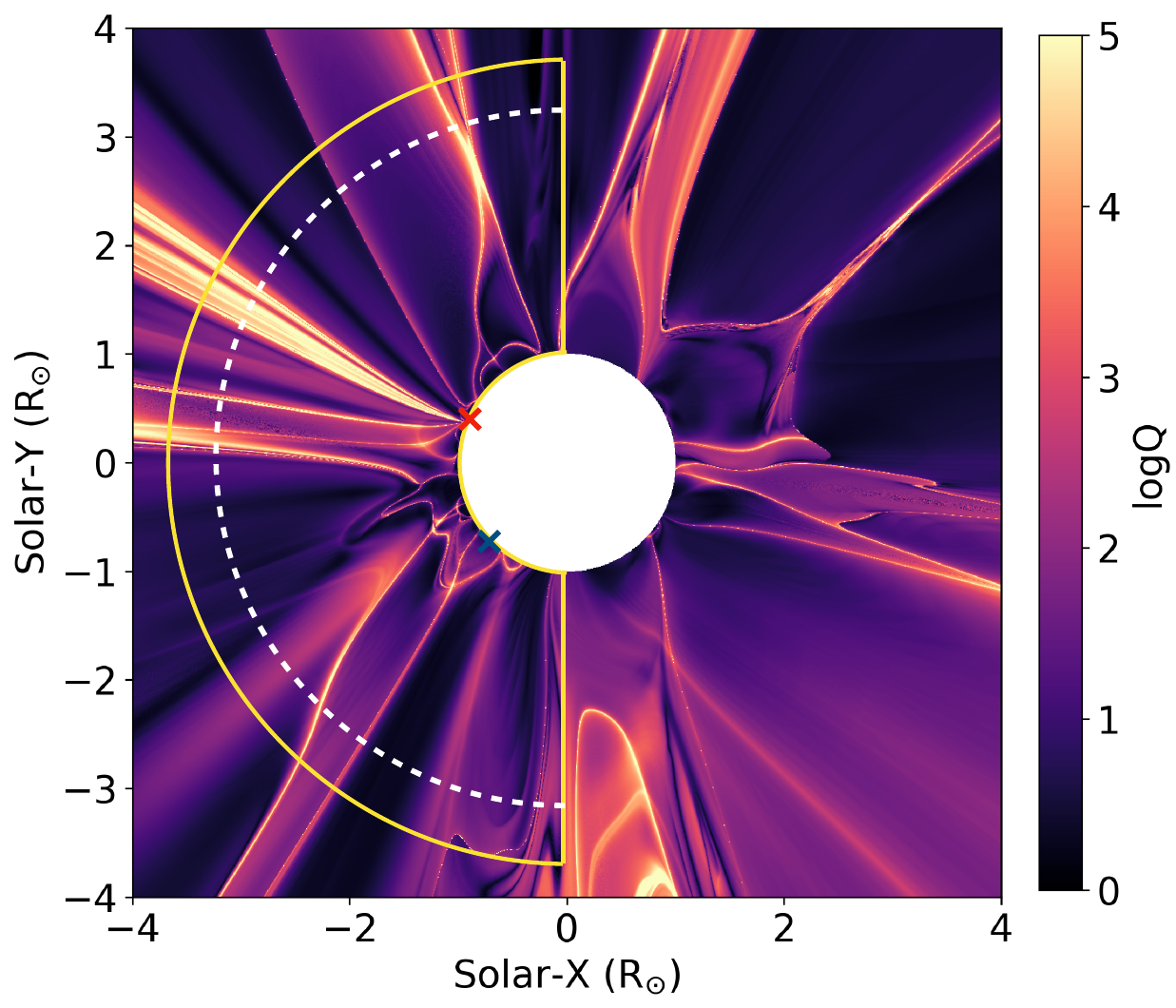}
    \caption{A map of squashing factor (logQ) values showing the magnetic structures of the low to middle corona as seen from SO on 2023 April 9. The red and blue crosses marked the location of two FOVs of EIS. The yellow contour illustrates the solar east limb region which corresponds to the polar projection in Figure \ref{fig:Fig9_PolarPlot}, and the white dashed line marks the heliocentric distance of 3~R$_\odot$.}
    \label{fig:Fig3_Qmaps}
\end{figure}

Another parameter called the squashing factor Q \citep{Titov2007}, is also computed to better characterise the boundaries between the open and closed magnetic field structures. In general, Q measures the deformation of circular magnetic flux tubes in the photosphere into elliptic flux tubes in the corona, which in turn quantifies the divergence of local magnetic field lines and gradients in field-line mapping. Hence, the Q value is large at the boundaries of two different magnetic topologies, such as at separatrix surfaces (Q $\rightarrow \infty$) or at quasi-separatrix layers (QSLs) \citep[Q \textgreater\ 10$^3$;][]{Antiochos2011}. The high Q regions, in which field lines from different magnetic domains converge, may be preferable sites for magnetic reconnection processes. In this analysis, we calculated Q from the coordinates and magnetic field values obtained from the results of the PSI-MAS simulation discussed earlier in this section. For each point on the surface, we defined flux tubes by tracing the field lines forwards and backwards 5 times (from the point itself and 4 neighbouring points). We then estimated Q from the coordinates and magnetic field properties at the boundaries of flux tubes \citep{mapfl}.

Figure \ref{fig:Fig3_Qmaps} shows the distribution of logQ values in the plane of the sky (POS) as seen from Metis during the observation period. The region above the EIS North FOV (red cross) consists of several complex high Q arcs (logQ \textgreater\ 3), particularly near the equatorial region. However, the region above the EIS South FOV generally has lower Q values with only a few high Q arcs. Hence, this further highlights the different magnetic environment in the middle corona above two different source regions of the solar wind.
    
\section{Low Corona Plasma Diagnostic - SDO and EIS}\label{sec:res1}
\subsection{Overview of the Solar Wind Source Regions}
\begin{figure*}[t!]
    \includegraphics[width=\textwidth]{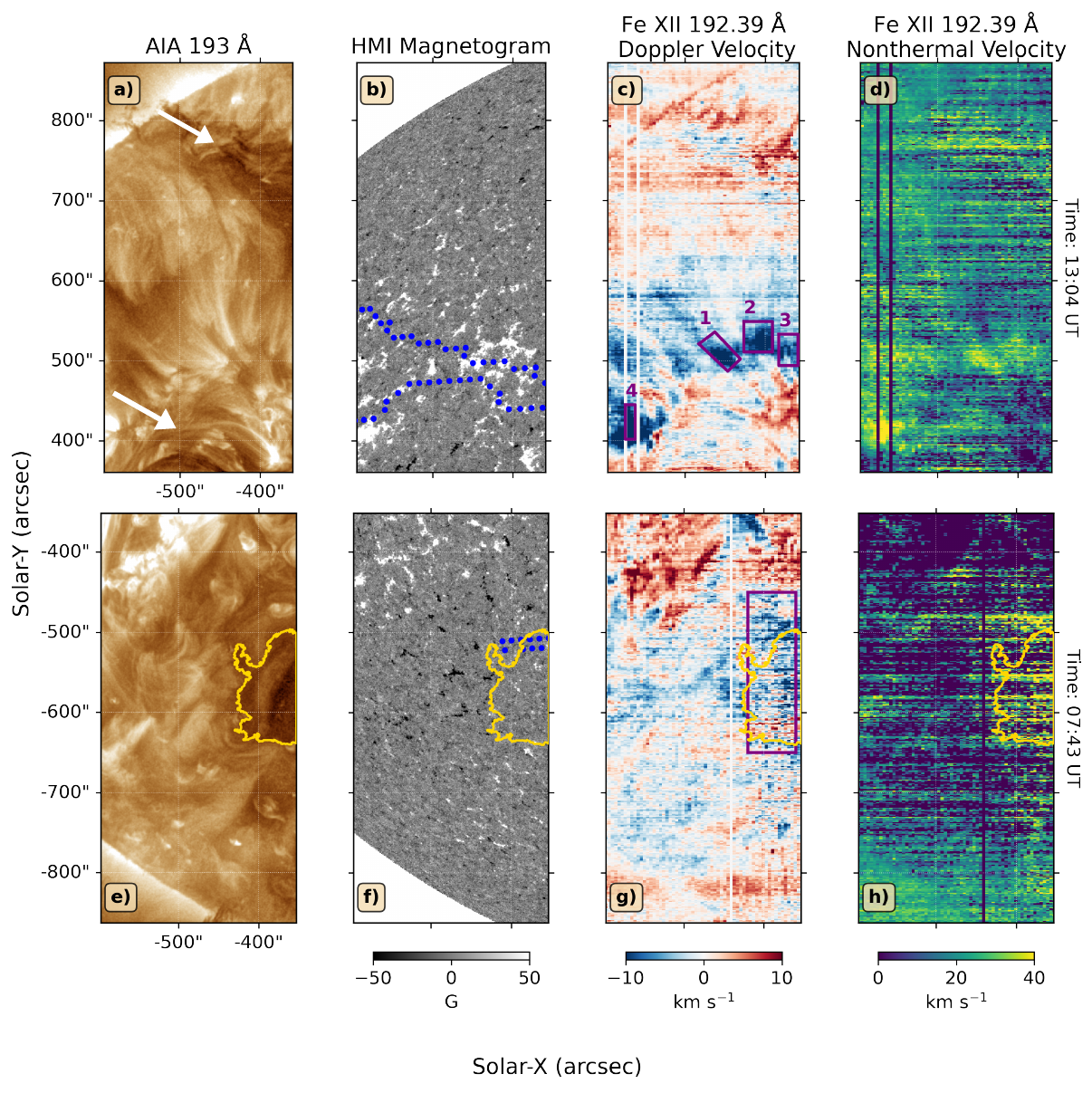}
    \caption{Plasma diagnostics of the low solar corona at the open-field corridor (EIS North FOV, top row) and CH (EIS South FOV, bottom row). From left to right, the columns show AIA 193~{\AA} observations, HMI LOS photospheric magnetograms, EIS Fe\,\textsc{xii} 192.39~{\AA} Doppler velocity maps, and EIS Fe\,\textsc{xii} 192.39~{\AA} nonthermal velocity maps. The yellow contours indicate the CH boundaries and the dotted blue contours outline the open-field locations.  The purple boxes indicate the regions where we average the spectra to obtain plasma properties shown in Table \ref{tab:Table1_EIS}. The colour scale of the Doppler velocity maps was set to [-10,10]~km~s$^{-1}$. The nonthermal velocity maps are saturated at [0,40]~km~s$^{-1}$. HMI magnetograms are saturated at $\pm$ 50~G. }
    \label{fig:Fig4_EIS}
\end{figure*}

The solar wind source regions in the low corona were observed by the AIA and HMI instruments onboard SDO and the EIS spectrometer onboard Hinode. Figure \ref{fig:Fig4_EIS} shows an overview of the low coronal observations at the EIS North FOV (top row) and at the South FOV (bottom row). Each row shows the AIA 193~\AA\ observations enhanced using the Multiscale Gaussian Normalisation technique \citep[MGN;][]{Morgan2014}, the HMI LOS magnetogram, and the Doppler velocity and nonthermal velocity maps obtained from fitting the EIS Fe\,\textsc{xii} 192.39~\AA\ spectral line. From panels a and e, the EIS North FOV corresponds to the decayed AR 12331 between two solar filaments (pointed by two white arrows), whereas the EIS South FOV corresponds to the eastern boundary of a small mid-latitude CH. The CH boundaries are defined using an intensity thresholding technique, with the threshold chosen to be 40\% of the median solar disc intensity in the 193~\AA\ passband \citep{Heinemann2019}. The boundaries are plotted in the yellow contours seen in the bottom row of Figure \ref{fig:Fig4_EIS}.

Panels b and f of Figure \ref{fig:Fig4_EIS} show the photospheric magnetic field strength in the northern and southern regions, with the location of the open field footpoints derived from the PSI-MAS model (see also Figure \ref{fig:Fig2_MAS}) denoted in dotted blue contours. This confirms that parts of the regions observed inside EIS's FOVs were magnetically open. Panel b also shows that the quiet Sun region observed in the EIS North FOV corresponds to a thin, latitudinal, open-field corridor at the base of the pseudostreamer (see panel a of Figure \ref{fig:Fig2_MAS}). Both footpoint locations correspond to positive-polarity magnetic field, with the open-field corridor being associated with stronger field than the CH counterpart. 

Note that the locations of open-field footpoints derived from the PSI-MAS model do not exactly match the locations of CH derived from EUV observations. The open-field corridor does not correspond to relatively dark regions (i.e., CH) on the solar disc. Instead, the corresponding region seems to consist of several bright plumes. The area of the CH in the EIS south FOV is also considerably larger than the open-field region derived from the model, which only captures the northern boundary of the CH. This mismatch may result from the limitation of magnetic modelling and photospheric magnetic field observations. Note that the disagreement between the area of open flux regions in the model and EUV observations is part of the so-called `Open Flux Problem' \citep{Linker2017, Asvestari2024}.

Doppler and nonthermal velocity maps can reveal the dynamics of plasma in the low corona. For the open-field corridor, plasma upflows, indicated by the blue-shifted region on the Doppler velocity map, were observed throughout the full longitudinal extent of EIS's FOV and spanning from 400\SI{}{\arcsecond} to 600\SI{}{\arcsecond} in Helioprojective latitude (see panel c of Figure \ref{fig:Fig4_EIS}). The upflow region appears to have a fan-like structure similar to the upflows seen at the edge of active regions \citep{Brooks2015, Yardley2021, Baker2023}. The base of the upflows has enhanced nonthermal velocities (see panel d), which also roughly correspond to the region with strong positive magnetic flux. The location of the upflows also coincides with the thin open-field corridor derived from the PSI-MAS model and the plume-like structures seen in the AIA 193~\AA\ passband.  Hence, our interpretation is that the upflowing plasma travels upwards along the open field lines, becoming the outflowing solar wind. 

We also found upflow regions within the southern CH and the surrounding boundary region. However, the upflow locations are more dispersed throughout the CH area (see panel g). There is also less enhancement in the nonthermal velocity compared to the north region. Since CHs are well known to have open magnetic field configurations, the plasma upflows inside CH are also thought to form part of the outflowing solar wind.

\subsection{Plasma and Magnetic Properties of Upflow Regions}
\begin{table*}[t!]
    \begin{tabularx}{\textwidth}{@{}lccccc@{}}
    \toprule
    \textbf{Location (Raster time)} & \multicolumn{4}{c}{Open-Field Corridor (13:04 UT)} &  CHB (07:43 UT)\\ \midrule
    \textbf{Box No.}                                                                                   & 1     & 2     & 3     & 4     & -    \\ \midrule
    \textbf{Mean Magnetic Flux Density (G)}          & 45.4  & 39.2  & 21.5  & 34.7  & 1.0  \\
    \textbf{Log$_{10}$ Density (cm$^{-3}$)} & 8.2   & 8.3  & 8.3  & 8.3  & 8.3 \\
    \textbf{FIP Bias}                                                                                  & 0.8   & 1.1   & 1.0   & 1.1   & 1.3  \\
    \textbf{Doppler Velocity (km s$^{-1}$)}                        & -6  & -14  & -9  & -8 & -5 \\
    \textbf{Nonthermal Velocity (km s$^{-1}$)}                     & 27 & 28 & 27 & 29 & 31 \\
    \textbf{Secondary Component}     & Yes & No & No & Yes & No \\          \bottomrule
     \end{tabularx}
\caption{Plasma and magnetic properties inside the regions of interest denoted by purple boxes in panel c and g in Figure \ref{fig:Fig4_EIS}. %The error of mean magnetic flux density is around 0.1 G
}
\label{tab:Table1_EIS}
\end{table*}
To quantitatively analyse the plasma parameters of these upflow regions, we defined several boxes based on the locations of strong upflows: four boxes in the EIS North FOV focusing on the regions with strong upflows in the open-field corridor and one box in the South FOV covering the eastern boundary of CH. The locations of these boxes are denoted as purple boxes in panels c and g of Figure \ref{fig:Fig4_EIS}. Plasma parameters were then derived using the averaged spectra of the pixels confined in each box to enhance the signal-to-noise ratio of the data \footnote{EIS Software Note: \url{https://zenodo.org/records/6339584}}. 

Table \ref{tab:Table1_EIS} shows the plasma and magnetic field measurements inside each box. Density measurements were derived using the spectral intensity ratio between Fe\,\textsc{xiii} 203.83~\AA\ and 202.04~\AA. The first ionisation potential (FIP) bias values were calculated using the diagnostic of the lines pair Si\,\textsc{x} 258.37~\AA\ -- S\,\textsc{x} 264.22~\AA\ following the method of \citet{Brooks2011}. The uncertainties are $\sim$30\% for log$_{10}$ density and FIP bias value, $\sim$5~km~s$^{-1}$ for Doppler velocity and $\sim$20~km~s$^{-1}$ for nonthermal velocity. Assuming the magnetic field is radial, we also reprojected the observed line-of-sight magnetic field component for each magnetogram pixel \citep{Hofmeister2017} before calculating the mean magnetic flux density inside each box.%should we mention issues with CH spectral lines?? (e.g., Young 2024, Wendeln and Landi 2018). 

In all boxes, we find that the plasma density and FIP bias values are very similar, with the density values in the range of $\approx$ 10$^{8.2}$ -- 10$^{8.3}$ and the FIP bias values between 0.8 and 1.3. These values correspond to those that typically characterise CH plasma, which has a relatively low density \citep{Hahn2010, Heinemann2021} and little to no enhancement of the low FIP elements \citep[i.e, FIP bias $\sim$ 1,][]{Feldman1998, Brooks2011}. 

The plasma dynamics in each box are also quite similar. For the open field corridor (Boxes 1 -- 4), the Doppler velocity ranges from -6 to -14~km~s$^{-1}$ and the nonthermal velocity ranges from 27 to 29~km~s$^{-1}$. For the CH region, the Doppler and nonthermal velocity values are -5~km~s$^{-1}$ and 31~km~s$^{-1}$. Note that the lower values may be the result of using relatively large macropixel boxes, which may average the strong upflows with surrounding weaker upflow regions.

The most striking differences between the open field corridor and the CH are the magnetic field properties. For the open-field corridor, the average magnetic flux density values range from 21 to 45 G, which are comparable to the values of decayed ARs \citep{Petrie2013} and the narrow AR upflow corridor \citep{Baker2023}. The mean magnetic flux density of the small CH is 1~G, which is similar to the typical values in nonpolar CHs \citep[$\approx$ 1 -- 5~G,][]{Hofmeister2017, Heinemann2019} and significantly lower than the values in the open-field corridor. 
\begin{figure*}[t!]
    \centering
    \includegraphics[width=\textwidth]{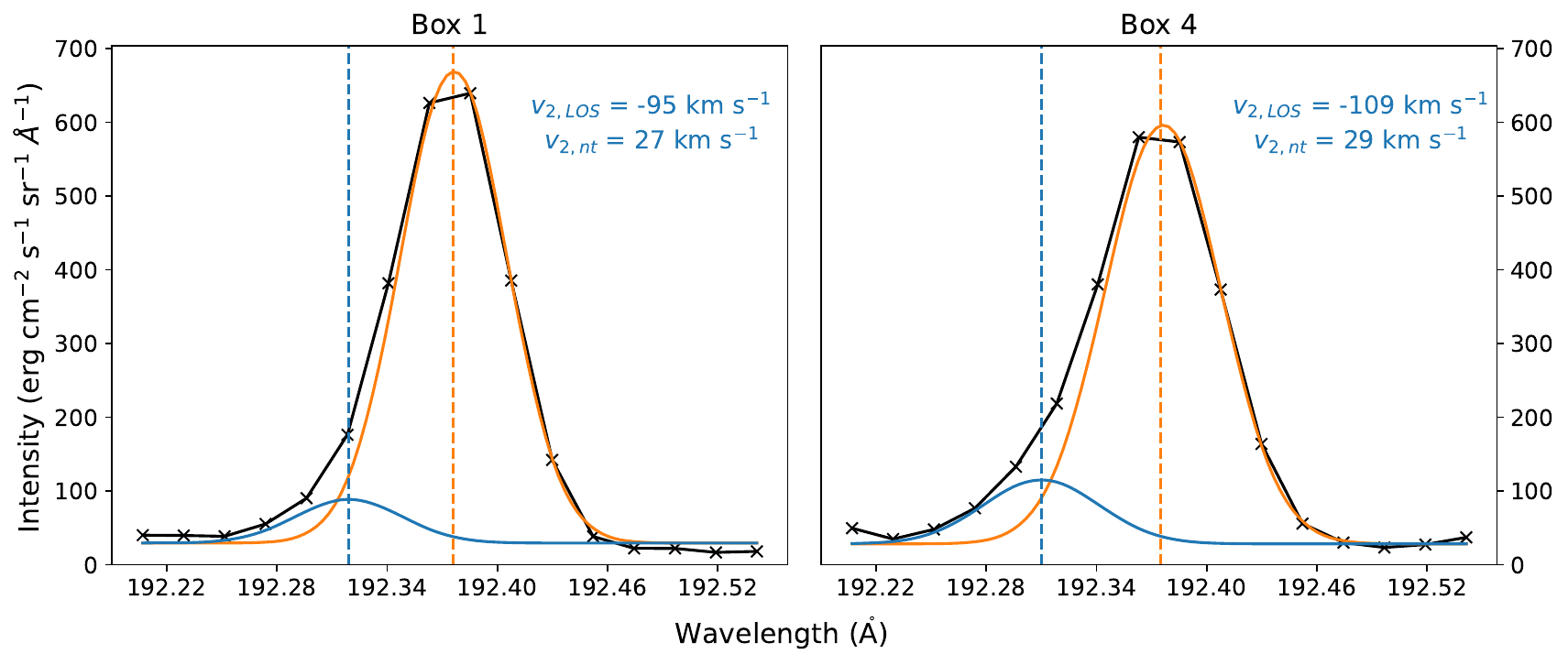}
    \caption{Fe\,\textsc{xii} 192.39~\AA\ averaged spectral line profiles obtained from Box 1 (left) and Box 4 (right), both correspond to the open-field corridor region. Both profiles can be fitted using the double Gaussian function, indicating the existence of the weak emission, high-speed secondary component of plasma upflow (blue). The calculated LOS velocity and nonthermal velocity of the secondary component are shown in each panel.}
    \label{fig:Fig5_Bluewing}
\end{figure*}

The averaged Fe\,\textsc{xii} 192.39~\AA\ spectral line profiles in Box 1 and Box 4 show significant blue-wing enhancements, suggesting that there is at least one secondary component of high-speed plasma upflows. To analyse these features, we used a double Gaussian function \citep{Brooks2012, Yardley2021, Ngampoopun2023} to fit both line profiles and derived the plasma dynamics of secondary upflows, as shown in Figure \ref{fig:Fig5_Bluewing}. In both profiles, significant secondary components are found, that is, the intensity of the secondary component is more than 10\% of the primary component. The LOS velocity of the secondary components is -95~km~s$^{-1}$ for Box 1 and -109~km~s$^{-1}$ for Box 4, which are comparable to the high-speed components of AR upflows \citep{Tian2021}. Since we choose to fit the double Gaussian with the same width, the nonthermal velocities of the secondary components are the same as their primary components counterpart, which is a reasonable assumption \citep{Tian2011}. These high-speed upflows are frequently observed near the edges of ARs \citep[e.g.,][]{Hara2008, Brooks2012, Yardley2021} and in coronal jets arising from open-field regions \citep{Young2014, Ngampoopun2023}.

\section{Middle Corona Plasma Diagnostic - Metis}\label{sec:res2} 
The properties of solar wind streams in the middle corona can be investigated using VL and UV coronagraph observations made by Metis. The two main properties that can be derived from these observations are the distributions of electron density and the outflow velocity of neutral hydrogen (H~\textsc{i}) in the corona.
\subsection{Electron Density} \label{subsec:el}
The observed polarised brightness (pB) in the VL passband mainly arises from Thomson scattering between photospheric photons and free electrons in the corona. Although light from interplanetary dust scattering may be slightly polarised \citep{Morgan2020, Boe2021}, pB emission from regions close to the Sun is still dominated by scattered light from electrons \citep[e.g.,][]{Lamy2020}. Therefore, the electron density in the middle corona can be determined on the basis of the pB distribution. The relationship between pB and electron density $n_e(r)$ is as follows \citep{vandeHulst1950, Hayes2001},
\begin{equation}
    pB = C\int_{x}^{\infty} n_e(r)[A(r) - B(r)]\frac{x^2 dr}{r\sqrt{r^2 - x^2}}
    \label{Eq1}
\end{equation}
where $A(r)$ and $B(r)$ are geometrical factors, $C = 3.44 \times 10^{-6}$~cm$^{-3}$ is a unit conversion factor, $x$ is the projected distance on the POS and $r$ is the heliocentric distance.

In this analysis, we follow the pB inversion method described by \citet{vandeHulst1950} and \citet{Hayes2001}. This method derives the electron density on the basis of two assumptions, that the distribution of electron density is axisymmetric along the axis of solar rotation, and the density profile can be expressed in the polynomial form
\begin{equation}
    n_e(r) = \sum_k (\alpha_k r^{-k})
\end{equation}
where $k$ is the polynomial degree and $\alpha_k$ is the unknown coefficient. By substituting the polynomial form into Equation \ref{Eq1}, $\alpha_k$ can be solved using the multivariate least-squares fitting method. For our calculation, we perform the fit for pB observations in the range of $r$ = 1.8 -- 3.5~R$_{\odot}$ to avoid noisy pixels close to the inner and outer FOV of the instrument. We also found that using $k$ = (1 -- 5) provided the best fitting results. The uncertainties are estimated to be approximately 10\% \citep[e.g.,][]{Dolei2018, Romoli2021}.

\begin{figure*}[t!]
    \includegraphics[width=\textwidth]{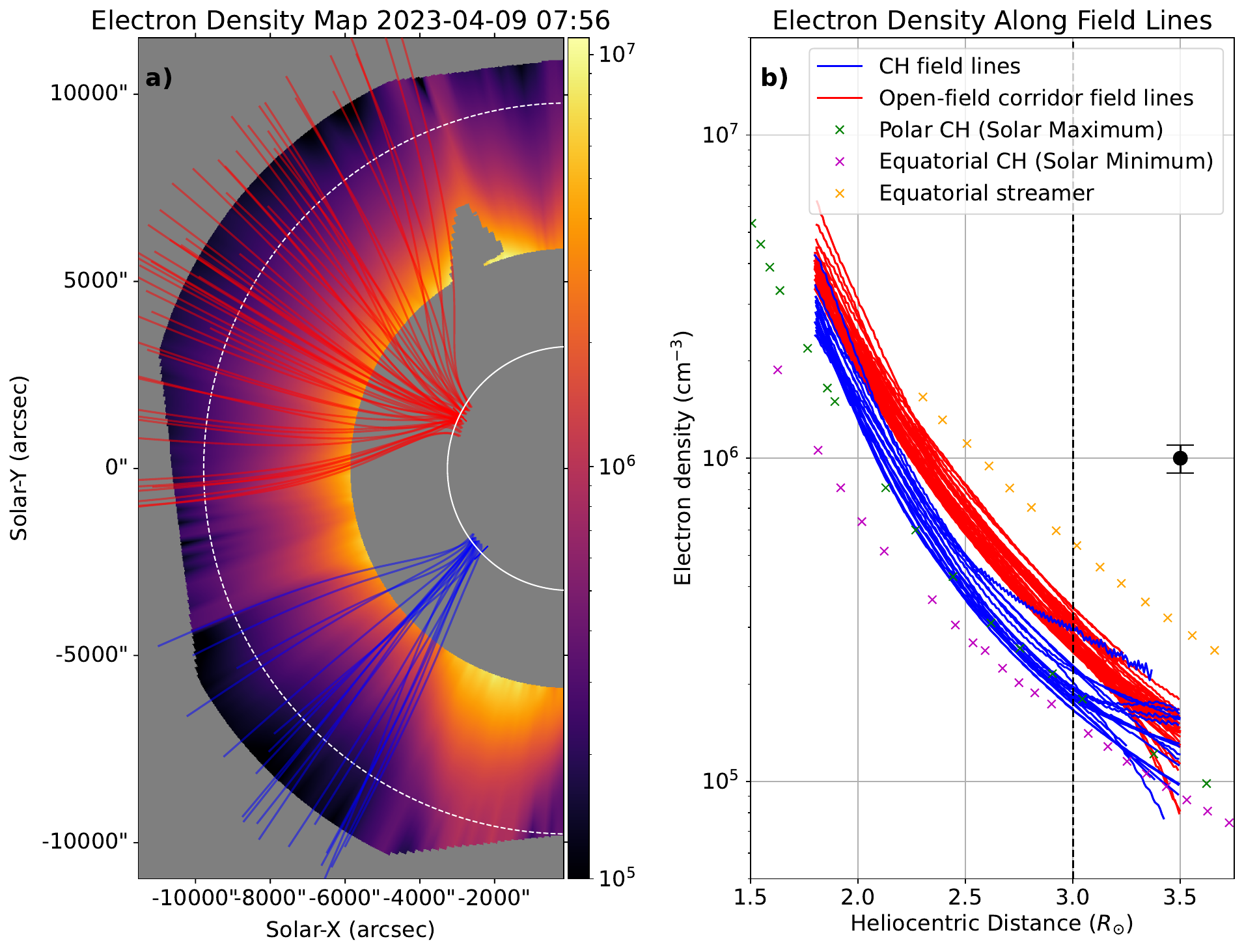}
    \caption{Electron density derived from pB inversion method. (a) Electron density map of the solar east limb seen by Metis, with overplotted extrapolated open field lines from the open-field corridor (red lines) and CH boundaries (blue lines). (b) Plot of electron density along the open field lines (red for open-field corridor and blue for CH boundaries) against distance from the Sun centre. The coloured crosses indicate the reported values of the electron density from equatorial CHs during solar minimum \citep[purple;][]{Withbroe1988}, polar CHs during solar maximum \citep[green;][]{Withbroe1988} and equatorial streamers \citep[orange;][]{Hayes2001}. The white and black dashed lines indicate the heliocentric distance $r$ = 3~R$_{\odot}$. The representative error of $\approx \pm10 \%$ is indicated as a black error bar.}
    \label{fig:Fig6_Ne}
\end{figure*}
Figure \ref{fig:Fig6_Ne} shows the electron density resulting from inversion of Metis VL pB observations at 07:56~UT. Panel a shows the 2D electron density map of the solar east limb as seen by Metis, which depicts the overall structure of the middle corona. During this period, the east limb consisted of a high-density streamer in the equatorial region and lower-density open-field regions surrounding it. The blue and red field lines denote the open fields rooted in the CH and the open-field corridor, respectively. The open field lines from both regions are derived from the PSI-MAS model as discussed in Section \ref{sec:model}.

The electron density values along these open field lines projected on the POS as a function of the heliocentric distance are displayed in panel b of Figure \ref{fig:Fig6_Ne}. In general, the electron density decreases by more than an order of magnitude as the heliocentric distance increases from 1.8 to 3.5~R$_{\odot}$. The electron density of the CH plasma (blue lines) is similar to the polar CH density at solar maximum and considerably higher than the equatorial CH density at solar minimum reported in \citet{Withbroe1988} (green and purple crosses). The electron density in the open-field corridor (red lines) is lower than the density in the equatorial streamers derived by \citet{Hayes2001} (orange crosses), but it is still noticeably higher than the CH plasma. The differences become more apparent at $r$ \textgreater\ 2.3~R$_{\odot}$, where the open-field corridor plasma has the density of $\approx 1.1 \times 10^6 \ \text{cm}^{-3}$ and the CH plasma has the density of $\approx 0.7 \times 10^6 \ \text{cm}^{-3}$. The differences also are visually evident on the electron density map. 

\subsection{{H~\textsc{I}} Outflow Velocity}
The UV Ly$\alpha$ intensity observed in the off-limb corona is mainly due to resonance scattering of chromospheric Ly$\alpha$ radiation by coronal H~\textsc{i} atoms \citep{Gabriel1971}. The coupling between protons and H~\textsc{i} atoms in the low to middle corona, caused by rapid charge transfers between them, allows us to use H~\textsc{i} atoms as a proxy for protons \citep{Allen1998, Kohl2006}. In the presence of the solar wind, the centroid of the incident chromospheric Ly$\alpha$ spectral line is Doppler-shifted from the centroid of the coronal H~\textsc{i} absorption profile, resulting in a systematic and progressive reduction in scattered intensity called the Doppler dimming effect \citep{Hyder1970, Withbroe1982,Noci1987}. The intensity is increasingly reduced with higher outflow speeds up to 450~km~s$^{-1}$ \citep[see Figure 1 in][]{Dolei2018}. Hence, we can exploit this effect to derive the outflow speed of H~\textsc{i} atoms (equivalent to proton speed) in solar wind streams. 

The derivation of coronal outflow speed based on the Doppler dimming effect involves creating the synthetic Ly$\alpha$ intensity based on several assumptions and input parameters. In particular, the outflow speed is treated as a free parameter in order to match the synthetic intensity with the observation made by the coronagraph. The method description and effects of the uncertainties of input parameters are extensively detailed in \citet{Dolei2018}.

Important input parameters include the electron temperature (T$_e$) and the hydrogen kinetic temperature (T$_{k}$), which cannot be directly determined from available coronal observations (from Metis or other instruments). Therefore, we have to adopt temperature profiles derived for similar coronal structures from previous analyses reported in the literature. In particular, \citet{Gibson1999} obtained the temperature profile from visible-light observations of large helmet streamers during the solar minimum, assuming the hydrostatic equilibrium condition. In addition, \citet{Vasquez2003} derived the analytical form of the temperature profile in the polar and equatorial regions, which is in good agreement with observations from UV coronagraphs. Both models are frequently used in the full Sun Doppler dimming analysis based on Metis observations \citep[e.g.,][]{Romoli2021, Antonucci2023}. Another critical assumption is the degree of anisotropy in T$_{k}$, defined as the ratio between kinetic temperature in direction perpendicular (T$_{k, \perp}$) and parallel (T$_{k, \parallel}$) to the magnetic field. Polar CH regions have been reported to show a strong anisotropy in T$_k$ \citep{Cranmer1999}, while equatorial regions typically show weaker anisotropy or isotropic conditions \citep{Vasquez2003, Spadaro2007}. 
% The anisotropy can lead to differences in the derived outflow velocity up to $\sim$ 30 -- 40~km~s$^{-1}$ \citep{Dolei2018, Antonucci2023}.

Since the coronal structures in our observations are quite complex as the Sun approaches solar maximum, we explored multiple cases using different temperature profiles relevant to both the equatorial and polar regions and different degrees of anisotropy to find reasonable assumptions that suit our observation. After careful consideration, we employed the following set of assumptions to create the synthetic Ly$\alpha$ intensity:
\begin{itemize}
    \item electron density derived from inversion of Metis pB images, as discussed in Section \ref{subsec:el}
    \item chromospheric Ly$\alpha$ line profile from analytical model by \citet{Auchere2005}
    \item uniform chromospheric Ly$\alpha$ line intensity across the solar disc with $I_\odot = 7.86\times10^{15}~\text{photon s}^{-1}~\text{cm}^{-2}~\text{sr}^{-1}$. The value is computed from the daily-averaged Ly$\alpha$ solar irradiance observed from Earth at the observation date, available at LASP Interactive Solar Irradiance Data Center\footnote{\url{https://lasp.colorado.edu/lisird/data/composite_lyman_alpha}}
    \item T$_e$ radial profile based on the profile of polar regions described in \citet{Vasquez2003}. The original profile is scaled so that the base temperature $\approx$ 1 MK, corresponding to the typical coronal temperature in the mid-latitude regions.
    % keep this item for future references if we use Gibson et al. 1999 model
    % \item uniform T$_{k, \perp}$ = 1.6~MK along radial direction \citep[see e.g.,][]{Romoli2021, Antonucci2023}. This value is approximately consistent with the T$_{k, \perp}$ calculated from Ly$\alpha$ spectral line width during solar maximum \citep{Dolei2016}, and also produces H~\textsc{I} spectral line profile with similar line broadening to those observed by Metis \citep{Antonucci2023}.
    \item mild anisotropy ($\approx$ 2) condition, with T$_{k, \parallel}$ = T$_e$ 
    \item T$_{k, \perp}$ radial profile based on the functional form given by \citet{Vasquez2003}. The profile is then adjusted to ensure mild anisotropy condition (i.e., T$_{k, \perp} \approx$ 2T$_e$) in the middle corona.
    \item integration of off-limb Ly$\alpha$ intensity along LOS of $\pm$ 10~R$_{\odot}$ from the plane of the sky.
\end{itemize}

To reduce some intensity fluctuations arising from an instrumental effect that affects the Metis UV detector \citep{Russano2024,Uslenghi2024,DeLeo2024}, we averaged 5 Ly$\alpha$ images acquired between 07:16~UT and 08:36~UT before deriving the coronal wind outflow velocity with the Doppler dimming technique. Note that the uncertainty in the derived speed due to calibration and data uncertainties is estimated to be on the order of $\sim 20~\text{km}~\text{s}^{-1}$ \citep[e.g.,][]{Antonucci2023}.

\begin{figure*}[t!]
    \includegraphics[width=\textwidth]{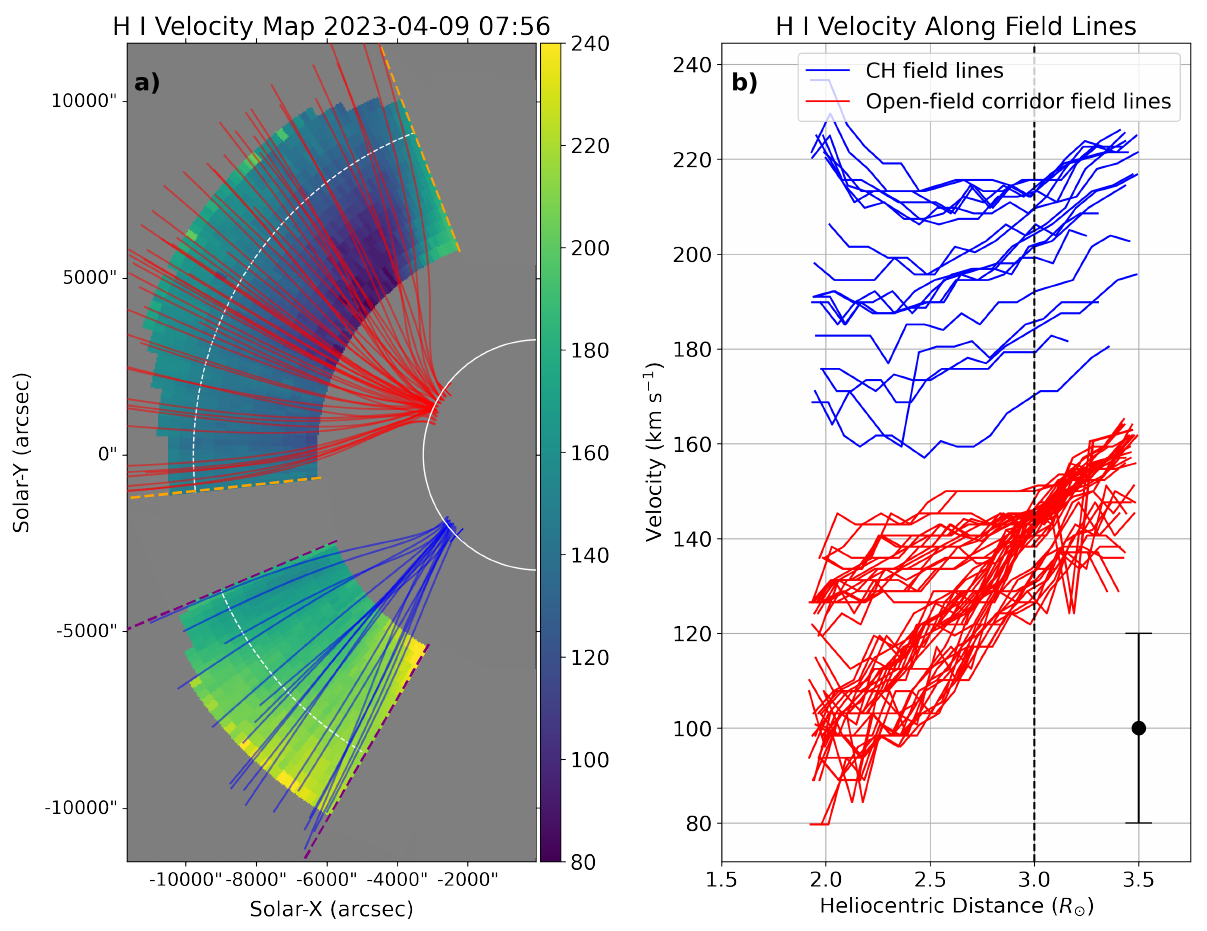}
    \caption{H~\textsc{i} outflow velocity derived from the Doppler dimming method. (a) H~\textsc{i} outflow velocity map of the solar east limb seen by Metis, with overplotted extrapolated open field lines from the open-field corridor (red lines) and CH boundaries (blue lines). Note that our assumptions apply only to regions bounded by orange dashed lines (for the open-field corridor) and purple dashed lines (for CH). (b) Plot of H~\textsc{i} outflow velocity values along the open field lines (red for the open-field corridor and blue for CH boundaries) against the heliocentric distance. The white and black dashed lines indicate a heliocentric distance of 3~R$_{\odot}$. The representative error of $\approx \pm20$ km s$^{-1}$ is indicated as a black error bar.}
    \label{fig:Fig7_vel}
\end{figure*}

Figure \ref{fig:Fig7_vel} shows the results of the solar wind velocity derivation using Doppler dimming analysis. Panel a displays the H~\textsc{i} outflow velocity map in the middle corona above the solar east limb seen by Metis, overplotted with the extrapolated field lines rooted in CH boundaries (blue) and the open field corridor (red), similar to Figure \ref{fig:Fig6_Ne}. The velocity values in the regions above the open-field corridor appear to be noticeably lower than those above the CH regions. Note that a single set of assumptions is insufficient to derive the outflow velocities from broad coronal regions with different characteristics. Hence, our assumptions (e.g., T$_e$, T$_{k}$, degrees of anisotropy) only apply to regions that correspond to the open-field corridor (bounded by orange dashed lines) and the CH (bounded by purple dashed lines), and they do not apply to other regions (coloured grey in Figure \ref{fig:Fig7_vel}), such as a helmet streamer near the south pole or an equatorial pseudostreamer.

The results are further highlighted in panel b where the velocity values along the open field lines projected on the POS are plotted against the heliocentric distance. It is evident that the solar wind streams from the CH have a consistently higher speed than those from the open-field corridor at all distances. The CH solar wind has a speed in the range of $\approx$ 160 -- 240~km~s$^{-1}$, while the open field corridor solar wind has a speed in the range of $\approx$ 80 -- 160~km~s$^{-1}$. The minimum speed of the CH solar wind at a low heliocentric distance ($\sim$ 160~km~s$^{-1}$) is comparable to the maximum speed of the open field corridor solar wind at the highest heliocentric distance. Both solar wind streams are moderately accelerated with increasing distance. However, the CH solar wind shows a peculiar flat or decreasing outflow velocity profile in regions below 2.5 R$_\odot$. We suspect that this might be due to the assumption of uniform chromospheric Ly$\alpha$ intensity, which leads to a considerable overestimation of the outflow speed in low-intensity CH at low coronal altitude \citep{Dolei2019}. This effect becomes less noticeable at larger heliocentric distances.
% apart from a few outliers in the CH solar wind, where the field lines are located close to the southern helmet streamer. 
% Interestingly, the outflow velocity profiles appear to be flatter in the CH compared to the open field corridor, indicating that the open field corridor solar wind has a higher acceleration rate with distance.   

\subsection{Correlation Between Plasma and Magnetic Field Properties in the Middle Corona}
\begin{figure}[t!]
    \centering
    \includegraphics[width=\textwidth]{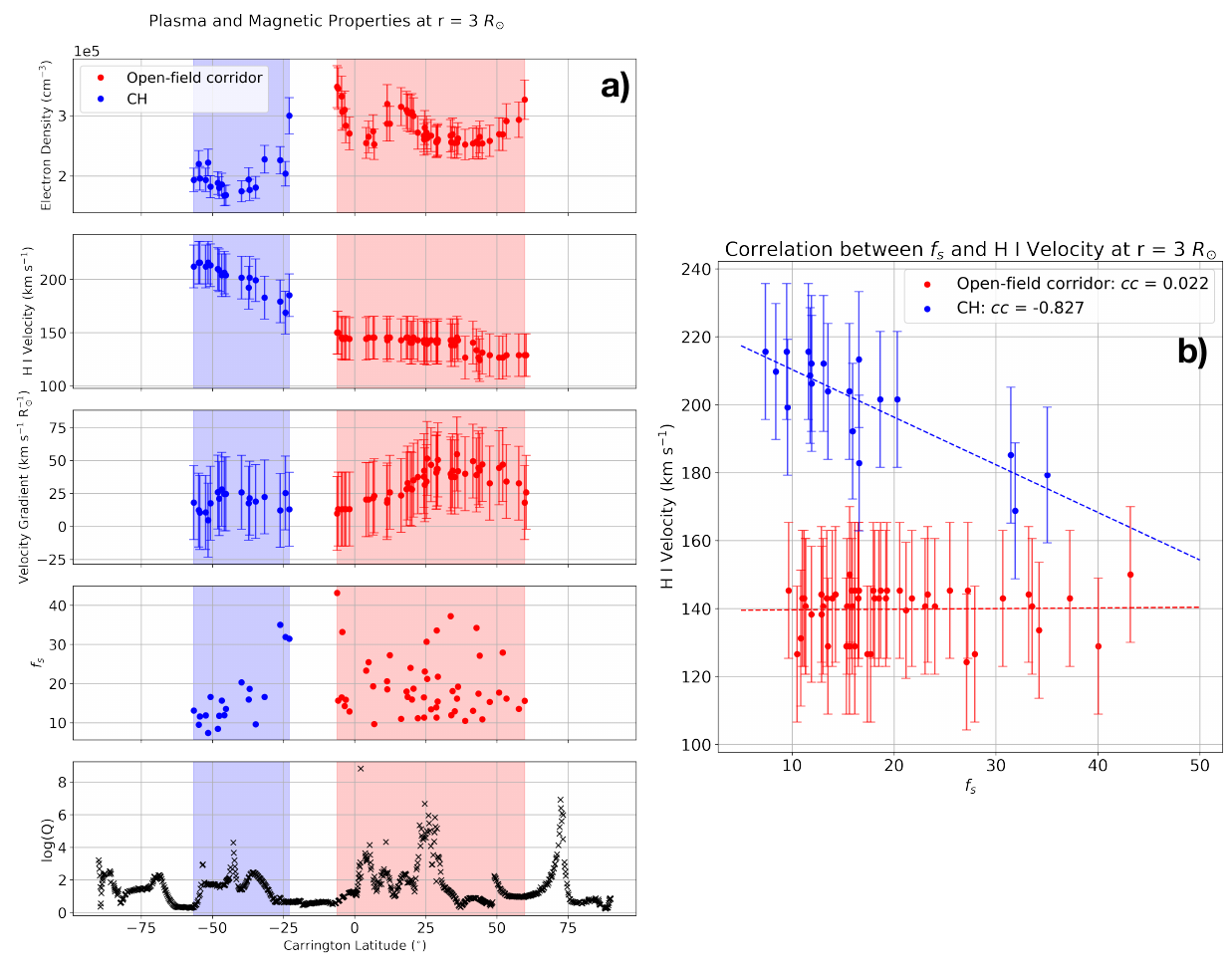}
    \caption{The plasma and magnetic field properties in the middle corona at a heliocentric distance of 3~R$_\odot$ above the solar east-limb seen by Metis. (a) Latitudinal distribution of electron density, H~\textsc{i} outflow velocity, velocity gradient (defined in text), $f_s$ and logQ. The blue (red) shaded region correspond to latitudinal extent of CH (open-field corridor) region. (b) The plot showing the correlation between H~\textsc{i} outflow velocity and $f_s$ for CH and open-field corridor solar wind. The Pearson correlation coefficient are displayed in the legend.}
    \label{fig:Fig8_PlasandMag}
\end{figure}
The electron density and outflow velocity maps, shown in Figures \ref{fig:Fig6_Ne} and \ref{fig:Fig7_vel}, allow us to directly compare solar wind streams arising from different coronal structures and investigate how their properties are distributed across the Carrington latitude. Figure \ref{fig:Fig8_PlasandMag} shows the plasma and magnetic properties in the middle corona above the solar east limb ($r$ =  3~R$_\odot$) against Carrington latitude. From panel a, we can clearly see the difference in the properties of the solar wind originating in two different source regions, with the CH solar wind (blue) having a lower electron density and higher H~\textsc{i} outflow speed compared to the open-field corridor solar wind (red). The average electron density in the CH is $\sim$ 1.5 times lower than in the open-field corridor, while the outflow speed in CH ranges from 170 to 210~km~s$^{-1}$ compared to $\approx$ 150~km~s$^{-1}$ in the open-field corridor. 

The average solar wind acceleration with distance is quantified as a velocity gradient, defined as $\Delta v/\Delta r$. We choose to calculate the average velocity gradient in the distance range $r=2.5-3.5~\text{R}_\odot$, to avoid the flat/decreasing velocity profile issue arising the lower altitudes of the CH.  The middle plot of panel a shows that the open-field corridor solar wind has a slightly higher velocity gradient than the CH solar wind, although the differences are within the estimated uncertainties.
% Similar results are reported in \citet{Antonucci2023}, in which they found that the slow solar wind near a helmet streamer has a higher acceleration than the faster wind surrounding it.

The expansion factors $f_s$ of the field lines arising from the CH and the open field corridor are generally of the same order of magnitude, with values ranging from $\sim$ 10 -- 40. Meanwhile, logQ values are higher in the open-field corridor compared to the CH (see also Figure \ref{fig:Fig3_Qmaps}), especially at Carrington latitude +25$^{\circ}$ where Q values reach 10$^6$.

The observed CH solar wind speeds, as inferred from the velocity of H~\textsc{i}, seem to have a clear trend of variation with Carrington latitude, with lower speeds near the equator and higher speeds closer to the south pole. This latitudinal variation is inverted in the distribution of $f_s$, with higher values near the equator and lower values closer to the pole. On the contrary, in the open-field corridor, the latitudinal distribution of the expansion factor $f_s$, appears to be higher and more scattered on average, without a specific latitudinal trend. This is reflected in the latitudinal distribution of the wind speed, which appears to be almost flat with small variations and values lower than in the CH region.

Panel b shows the correlation between the solar wind speed and the expansion factor $f_s$ in the middle corona. We find that the solar wind speed is inversely correlated with the values of $f_s$, with the Pearson correlation coefficient ($cc$) of -0.827. This inverse correlation is in line with the empirical relationship between the expansion factor and the solar wind speed at 1 AU \citep{Wang1990}. However, we find no correlation between solar wind speed and $f_s$ in the open-field corridor, with a $cc$ of 0.022. 
% There is also no significant correlation between solar wind speed and Q in both the CH and the open-field corridor.

\section{Dynamics In the Low-Middle Corona} \label{sec:res3}
\begin{figure*}[t!]
    \centering
    \includegraphics[width=0.85\textwidth]{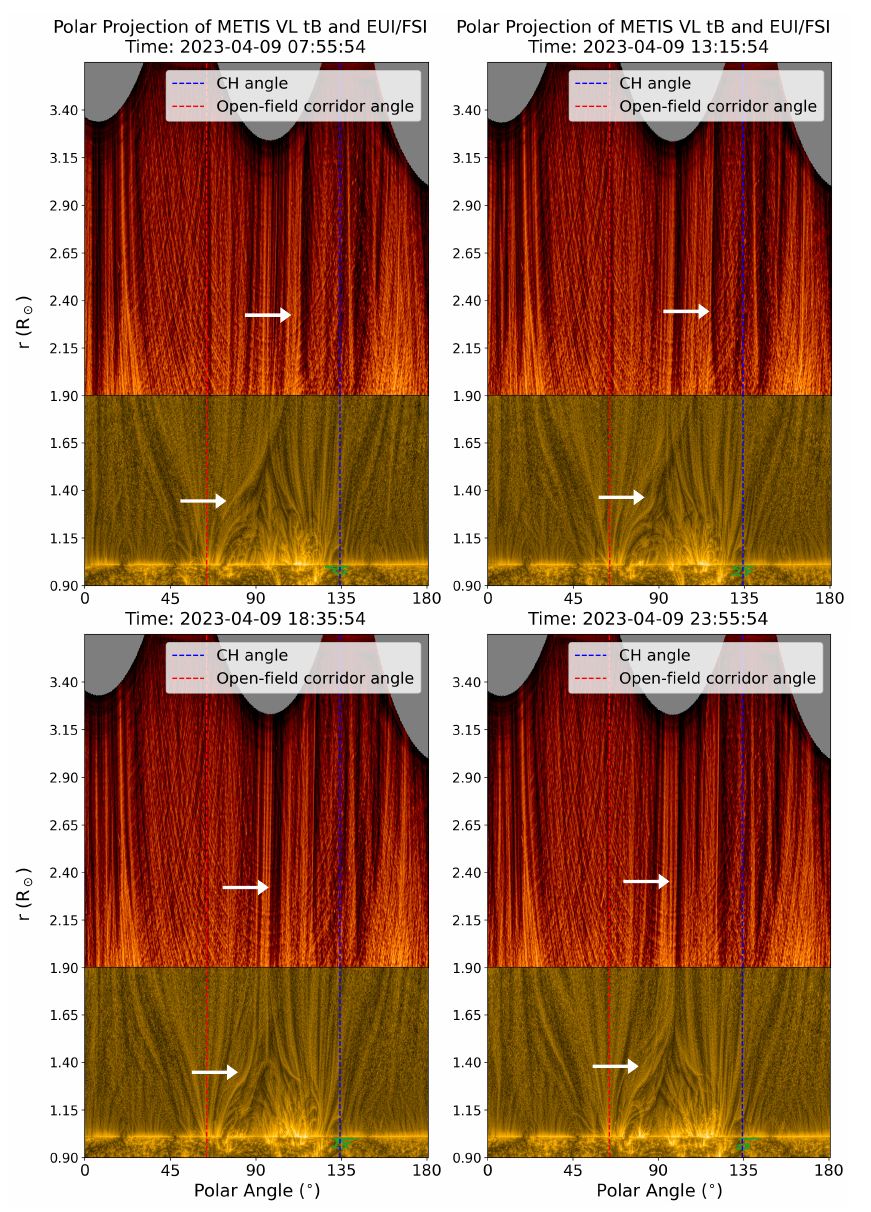}
    \caption{Sequence of the polar projection of combined EUI/FSI 174 {\AA} and Metis VL tB observations of the solar east limb from a heliocentric distance of 0.90 to 3.65~R$_{\odot}$. The red and blue dashed lines indicate the approximate polar angle of the open-field corridor and CH, respectively. The green contour marks the boundary of the CH. The white arrows are plotted to guide the eye to see the evolving boundaries of the open-field corridor in the low and middle corona. An animated version of this figure is available as Figure9Animation.mp4. The movie has a duration of 11~s and shows the evolution of the solar east limb observed from EUI/FSI and Metis from 04:56~UT to 23:56~UT.}
    \label{fig:Fig9_PolarPlot}
\end{figure*}

Fine-scale structures in the middle corona are difficult to observe in the EUV passbands, especially in the open-field regions with low density. EUV emission mainly arises from collisional processes with its intensity proportional to the density squared ($I \propto n_e^2$), which means that the EUV intensity decreases rapidly as the electron density decreases with heliocentric distance. This effect is lessened in visible-light coronagraph observation, in which the intensity depends on density ($I \propto n_e$, see also Equation \ref{Eq1}). However, the limitations of previous space-based coronagraphs leave the regions below a heliocentric distance of $\sim$2 R$_\odot$ relatively underexplored. Since the inner FOV of Metis and the outer FOV of EUI/FSI partially overlap, it is possible to seamlessly study the evolution of coronal structures from near the solar surface up to the middle corona with high spatiotemporal resolution EUV and visible light observations during the Solar Orbiter's perihelion.

Figure \ref{fig:Fig9_PolarPlot} shows snapshots of the solar east limb observed by EUI/FSI 174~\AA\ and Metis VL tB at three different times during the observation date. Both observations are enhanced using the wavelet-optimised whitening method \citep[WOW;][]{Auchere2023} and projected into the polar coordinate system, where the polar angle starts from the solar north pole (0$^\circ$) and goes counterclockwise (see the yellow contour in Figure \ref{fig:Fig3_Qmaps}). The combined FOV of the composite observation covers the heliocentric distance $r$ = 0.90 -- 3.65~R$_\odot$, with SO/EUI showing the low corona in EUV and Metis showing the middle corona in visible light. The animated version of Figure \ref{fig:Fig9_PolarPlot} shows the composite SO/EUI-Metis observations of the off-limb corona above the east limb from 04:56~UT to 23:56~UT, with a cadence of 20~min. 

These combined observations reveal the highly structured and extremely dynamic nature of the low and middle corona. The structures seen in the EUV and VL passbands can be smoothly connected at the height of the boundary between the two instrument FOVs ($r$ = 1.9~R$_\odot$). The evolution in each fine structure seen in the VL passband in the middle corona can also be traced back to the corresponding evolving EUV structures in the low corona.

There is an extended streamer near both the north and the south poles. Both streamers can be identified as the region comprised of low-emission EUV loops and brighter visible light strands in the middle corona. The equatorial pseudostreamer (polar angle $\sim$ 90$^\circ$), on the other hand, consists of numerous bright EUV loops, which may result from the embedded AR (see the left panel in Figure \ref{fig:Fig1_AIAandSO}) inside the pseudostreamer. The pseudostreamer cusp is located inside the EUI/FSI FOV at $r$ = 1.6~R$_\odot$, which is lower than polar streamer cusps ($r$ \textgreater\ 2~R$_\odot$). The angular width of the pseudostreamer is well confined between the open-field corridor (red dashed line, polar angle $\sim$ 65$^\circ$) and the CH (blue dashed line, polar angle $\sim$ 135$^\circ$)

The open-field corridor, located to the left (north) from the pseudostreamer, is funnel-shaped and filled with numerous strands of plasma. These strands show persistent upward flows, indicating that the plasma is outflowing from the low corona to higher altitudes. The angular extent of the funnel continuously expands with heliocentric distance, tapering the shape of the north pole streamer and the equatorial pseudostreamer. For the CH on the right (south) side of the pseudostreamer, we also identify a funnel-shaped plasma outflow similar to the open-field corridor. We interpret these outflowing plasma strands as the tracers of the solar wind. However, the bright strands seen in EUV and VL may not necessarily arise from the CH, as there is another open field region at the edge of the AR located behind the limb (see panel a of Figure \ref{fig:Fig2_MAS}). Hence, it may be possible that those bright strands are from behind the limb and can be seen because of the LOS integration effect. 
%add discussion on ray-like structure

The elongated ray-like strands correspond to fine-scale structures in the open-field regions, commonly referred to as plumes \citep{Poletto2015} or plumelets \citep{Uritsky2021, Morton2023}. These structures may extend from the low corona up to a distance of 45 R$_{\odot}$ \citep[e.g.,][]{DeForest2001} and highlight the nonuniformity of plasma and magnetic field structures in the open-field region \citep{Boe2020}. Several twisted or helical structures, previously seen in total solar eclipse images \citep{Druckmuller2014} can be identified near the pseudostreamer boundary in EUV observation, which may also indicate the presence of plasma instabilities.

The animated version of Figure \ref{fig:Fig9_PolarPlot} also shows the continuous reconfiguration of the structures in the low and middle corona. The reconfiguration of the structures can be seen as the formation of new loops or the deformation/brightening of existing structures in EUV, or the positional shifts in visible light strands above 1.9~R$_\odot$. Since the plasma is frozen into the magnetic field in the corona, the reconfiguration of these structures can be inferred as tracers of the evolution of magnetic field lines due to magnetic reconnection processes. The white arrows in Figure \ref{fig:Fig9_PolarPlot} point toward an example of reconfiguration of the coronal structure, which is the boundary of the open-field corridor and an equatorial pseudostreamer. We observe a clear evolution of the boundary in the EUI/FSI passband over four time steps shown in Figure~\ref{fig:Fig9_PolarPlot}, where the shape of the boundary is significantly changed. Although the evolution is less evident in the Metis observations, we can still observe the persistent shift of the boundary (pointed by arrows) toward a lower polar angle (higher latitude). The movie version of Figure~\ref{fig:Fig9_PolarPlot} reveals that these events can be ubiquitously found throughout the observation period, and they seem to be more evident at the boundaries between the plasma outflow funnel and the streamer.

\section{Discussion and Conclusion} \label{sec:disc}
The coordinated observations between Solar Orbiter, SDO, and Hinode allow us to identify and analyse solar wind streams emanating from two different source regions: a mid-latitude coronal hole and an open-field corridor. The solar wind plasma properties are investigated in the low corona using EIS and in the middle corona using Metis. Magnetic field extrapolations based on MHD modelling are used to verify the connection between solar wind sources in the low and middle corona and to provide the global magnetic field configuration, which plays an important role in solar wind formation and acceleration.

By comparing the solar wind plasma properties (electron density and outflow velocity) from the two sources at two different solar altitudes, we find that the differences between the two solar wind streams are more pronounced in the middle corona. Plasma density and composition (as inferred by FIP bias) are very similar for both the CH and the open-field corridor solar wind, while their plasma dynamics (i.e. Doppler and nonthermal velocity) are slightly different (see Table \ref{tab:Table1_EIS}). However, in the middle corona, we can clearly distinguish these two solar wind streams, with the open-field corridor solar wind generally having higher electron densities and lower outflow speeds than the CH solar wind (see Figure \ref{fig:Fig8_PlasandMag}).

Note that the plasma properties in the low and middle corona are derived from different populations of plasma. This is particularly important for the derivation of the outflow velocity, which represents the dynamics of Fe ions in the low corona and of neutral hydrogen atoms in the middle corona. Although both elements can be interpreted as solar wind outflows, directly linking their dynamics is nontrivial. The feasibility of connecting the Doppler velocity from spectroscopy with the H~\textsc{i} velocity from the Doppler dimming technique is interesting and can be explored in future work.
%Discuss the limitation and caveats of Metis and EIS

We acknowledge that the complex structure of the corona near solar maximum makes the interpretation of the data challenging. One of the main limitations of this work is that although we establish the connection between EIS and Metis through a magnetic field extrapolation, it is difficult to confirm whether Metis was observing the same features as EIS. This difficulty arises from the limited FOV of EIS and also from the LOS integration effect in off-limb observations. The superimposed structures along the LOS may also considerably affect the derivation of electron density and outflow velocity. However, our results for the electron density of the CH agree with previous observations of CH at solar maximum \citep{Withbroe1988}. Note that most of the CH electron density values reported in the previous literature \citep[e.g.,][]{Guhathakurta1999, Hayes2001, Morgan2007} are lower than our values because the measurements take place in different phases of a different solar cycle \citep{Ventura2005, Antonucci2020b}. The electron density values for the open-field corridor also seem to be reasonable since they are between the values of the CHs and the equatorial streamers \citep[see also][]{Abbo2010}. The LOS integration effect is also taken into account in the assumption for the Doppler dimming technique used to derive H~\textsc{i} outflow velocities. Hence, our results are still valid despite these caveats.

Another limitation is that we do not have simultaneous measurements of electron and hydrogen kinetic temperature in the middle corona that complement Metis observations. This problem could be alleviated in the future by new instruments, such as the Coronal Diagnostic Experiment \citep[CODEX;][]{Casti2024, Gong2024}, which is capable of measuring coronal temperature in the heliocentric distance range of 3--10 R$_\odot$.

\subsection{Nature of Open-Field Corridor}
Despite the similarity in plasma composition and density, the open-field corridor and CH can still be distinguished by their differences in appearance in the EUV corona and magnetic-field properties. The magnetic flux density of the open-field corridor in the photosphere is in the order of 20 -- 50~G, which is significantly higher than the CH counterpart. This disparity implies that, although both regions host an open magnetic field configuration, the magnetic environments of the open-field corridor and CHs (including surrounding boundary regions) are fundamentally different.

We find that the open-field corridor corresponds to a decayed AR, which is commonly associated with the formation of nonpolar CH \citep{Karachik2010, Petrie2013}. The corridor is not obviously darker than the surrounding region in AIA 193~\AA, which could be the result of the LOS effect of several bright plumes (see panel a in Figure~\ref{fig:Fig4_EIS}) and the remnants of AR loops that outshine the dark regions adjacent to them. The brighter EUV emission of the open-field corridor compared to the CH, even though both regions have a similar electron density (see Table \ref{tab:Table1_EIS}), may also imply that it has a somewhat higher electron temperature. However, since the derived density is taken from measurements at CH boundary regions, the actual CH density may also be slightly lower than that of the open-field corridor, resulting in lower emission. The higher electron temperature and stronger magnetic field in the open-field corridor suggest that there are additional heating mechanisms that may affect the formation of solar wind emanating from it.

\citet{Wang2019} found that some equatorial CHs during the solar maximum may not appear dark because of nearby bright loops from a nearby AR or its remnant. They also find a magnetic field strength of the order of 30~G and a high expansion factor ($f_s$ \textgreater\ 9). The properties of the open-field corridor in our observation are comparable to this result, suggesting that it could be categorised as a thin mid-latitude CH. 

However, the open-field corridor can also be compared with the dark channel near an AR observed by \citet{Baker2023}. In particular, the plasma density, plasma dynamics, and magnetic-field properties inside the open-field corridor are approximately consistent with their analysis. Note that the FIP bias in the \citet{Baker2023} channel is higher than in our observation, probably due to close proximity to an AR. \citet{Baker2023} interpret their dark channel as the narrow open-field corridor associated with the S-web model, due to its extremely high expansion factor (up to 300), topological robustness, association with a pseudostreamer, and also the very low solar wind speeds observed in situ. These characteristics are also applicable to our observations, suggesting that the open-field corridor could be associated with the S-web structure in the solar corona. The high values of log Q above that region (see Figure \ref{fig:Fig3_Qmaps}) also support this argument, as they indicate the existence of separatrix surfaces and QSLs. 

Therefore, we propose that this open-field corridor is a narrow mid-latitude coronal hole that forms part of the S-web structure due to its high squashing factor and the association with the leg of a pseudostreamer. Note that S-web is defined purely based on magnetic modelling, and it can encompass various structures in the low corona, including both CH and AR \citep{Chitta2023b, Baker2023}. 
% add discussion on open-flux problem and fine-scale structure here.

The open-field corridor also serves as direct evidence that not all open-field lines originate from CHs traditionally defined based on EUV observations. \citet{Asvestari2024} shows that the open field regions derived from various coronal models (including PSI-MAS) are considerably mismatched with CH area extracted from EUV image, even for during solar minimum, where CH is better observed. We suspect that the mismatch will become even more evident as the solar cycle progress toward maximum, as shown from our analysis in which the open-field corridor does not appear dark in AIA 193 \AA\ . Moreover, \citet{Boe2020} shows that open field lines seem to be abundant in total solar eclipse observations even without the presence of CHs on the solar disc. This mismatch between modelling and observations needs to be further addressed in order to resolve the open flux problem \citep{Linker2017}.

\subsection{Reconnection Dynamics Driving Solar Wind Variability}
By using the visible-light pB inversion and Doppler dimming technique to analyse the data from Metis, we can derive the solar wind electron density and outflow velocity map in the middle corona and can directly compare the distribution of solar wind properties emanating from the CH and the open-field corridor. We obtain two key results from the analysis. First, we find that the electron density and solar wind velocity arising from two different sources are distinctively different in the middle corona compared to the low corona, suggesting that certain processes arise between the low and middle corona that drive these differences. Second, we also find that the speed of the solar wind from the open-field corridor does not appear to be inversely correlated with the expansion factor in the middle corona ($cc$ = 0.022), unlike the solar wind from the CH ($cc$ = -0.827). In the expansion factor framework, the solar wind heating distribution and subsequently the solar wind speed depend on the geometry of flux tubes. Therefore, this non-correlation between solar wind speed and expansion factor suggests that there should be other phenomena that lessen the effects of flux tube geometry on solar wind speed. 

The structure of the low-middle corona is complex and dynamic, as suggested by the coobservation of EUI/FSI and Metis (Figure \ref{fig:Fig9_PolarPlot} and accompanying animation) and the calculation of the squashing factor logQ (Figure~\ref{fig:Fig3_Qmaps}), which serves as a proxy for the complex magnetic environment preferred for reconnection. By comparing Figure~\ref{fig:Fig3_Qmaps} and Figure~\ref{fig:Fig9_PolarPlot}, it is evident that the boundary between the equatorial pseudostreamer and the expanding funnel of the open-field corridor approximately corresponds to high logQ regions, while the CH region generally has a lower logQ (see also Figure~\ref{fig:Fig8_PlasandMag}). Therefore, we can also infer that reconnection may occur more readily inside the open-field corridor and its boundary than in the CH because of the more complex magnetic field structure.

Numerous cases of plasma structure reconfiguration observed inside or nearby open-field regions indicate that the solar wind outflow is not steady and smooth but rather is constantly undergoing the process of magnetic reconnection, especially along the boundaries of different magnetic domains (e.g., the edge of pseudostreamer), as indicated by the high logQ arcs. Therefore, we hypothesise that these ubiquitous reconnection processes play an important role in driving the variability of solar wind streams.

\citet{Chitta2023b} identified highly structured elongated features in EUV observations of the middle corona, which they called the `coronal web'. In their observations, the coronal web seemed to emanate from a large region of complex magnetic structure with a high squashing factor, and each individual coronal web continuously undergoes reconfiguration. Hence, they interpreted the coronal web as the direct imprint of S-web dynamics in the middle corona that drives the slow solar wind through magnetic reconnection. Our observation of structured plasma outflow strands in EUV and VL images could correspond to similar coronal web features reported in \citet{Chitta2023b}, as both structures are similar in dynamics, spatial extent, and association with high squashing factors. In particular, in the open-field corridor, we notice that the reconfiguration events are more evident and the squashing factor values are considerably higher compared to the CH region. This might imply that there are numerous solar wind streams generated through reconnection events in the low and middle corona in our observations. 

The solar wind streams generated in this way will likely have different properties compared to the solar wind originating from the sources in the low corona (i.e. upflow regions). For example, the plasma density is likely to be intermediate between the values in the two reconnecting structures, and the speed of plasma may also be related to the energetics of each reconnection event itself rather than the expansion factor of flux tubes. These newly generated solar wind streams will inevitably mix with existing streams, and this might help explain why the difference in properties of solar wind from between the open-field corridor and the CH is more pronounced in the middle corona.

Therefore, our work highlights the importance of magnetic reconnection for solar wind formation and acceleration, in line with the S-web model. The highly dynamic nature of the middle corona and the increasingly complex magnetic field structures as the solar cycle progresses towards the maximum serve as the ideal environment for reconnection to occur more ubiquitously and to directly impact the variability of solar wind. However, the intrinsic properties of the source regions, such as magnetic field strength and fine-scale structures inside open-field regions, are still important for solar wind variability and cannot be completely ruled out. Our work is also in line with the recent framework introduced by \citet{Viall2020}, which challenges the historical paradigm of solar wind bimodality by suggesting that the solar wind can go through multiple pathways involving different processes, resulting in various types of solar wind parcels in contrast with the traditional fast-slow distinction.

Finally, this research also demonstrates the ability to combine spectroscopic and coronagraph observations from different vantage points, which proves to be powerful in helping us better understand the origin of the solar wind. 

%%%%%%%%%%%%%%%%%%%%%%%%%%%%%%%%%%%%%%%%%%%%%%%%%%%%%%%%%%%%%%%%%%%%%%%%%%%
\begin{acks}
 The authors wish to thank the anonymous referee who provides insightful and constructive comments that improve the manuscript. The authors also thank Lakshmi Pradeep Chitta for his helpful discussions that improved the content of this article. Solar Orbiter is a space mission of international collaboration between ESA and NASA, operated by ESA. Metis was built and operated with funding from the Italian Space Agency (ASI), under contracts to the National Institute of Astrophysics (INAF) and industrial partners. Metis was built with hardware contributions from Germany (Bundesministerium für Wirtschaft und Energie through DLR), from the Czech Republic (PRODEX), and from ESA. The authors thank the former Metis Principal Investigator, Ester Antonucci, for leading the development of Metis until the final delivery to ESA. The EUI instrument was built by CSL, IAS, MPS, MSSL/UCL, PMOD/WRC, ROB, LCF/IO with funding from the Belgian Federal Science Policy Office (BELSPO/PRODEX PEA 4000134088); the Centre National d’Etudes Spatiales (CNES); the UK Space Agency (UKSA); the Bundesministerium f\"{u}r Wirtschaft und Energie (BMWi) through the Deutsches Zentrum f\"{u}r Luftund Raumfahrt (DLR); and the Swiss Space Office (SSO). Hinode is a Japanese mission developed and launched by ISAS/JAXA, collaborating with NAOJ as a domestic partner, NASA and STFC (UK) as international partners. Scientific operation of the Hinode mission is conducted by the Hinode science team organized at ISAS/JAXA. This team mainly consists of scientists from institutes in the partner countries. Support for the post-launch operation is provided by JAXA and NAOJ (Japan), STFC (UK), NASA (USA), ESA, and NSC (Norway). SDO is a mission of NASA's Living with a Star programme. SDO data are courtesy of NASA/SDO and the AIA, EVE, and HMI science teams. This research made use of several open-source Python packages including NumPy \citep{Harris2020}, Astropy \citep{AstropyCollaboration2013}, Matplotlib \citep{Hunter2007}, SciPy \citep{Virtanen2020}, SunPy \citep{SunPyCommunity2020}, EISPAC \citep{Weberg2023}, and aiapy \citep{Barnes2020}. The IDL software used in this work used the SSW packages \citep{Freeland1998}.

\end{acks}

% \noindent To change a title use an optional parameter:\par
% \verb+\begin{acks}[Acknowledgements]...\end{acks}+

%\acknowledgment US spelling: \verb+\acknowledgment+
%\acknowledgement British  spelling: \verb+\acknowledgement+

%%%%%%%%%%%%%%%%%%%%%%%%%%%%%%%%%%%%%%%%%%%%%%%%%%%%%%%%%%%%%%%%%%%%%%%%%%%
\begin{authorcontribution}
N.N. led the project,  performed most of the data analysis, prepared all figures, and wrote the manuscript. R.S., D.H.B., L.A., D.S. and D.B. contributed to the project conceptualisation. R.S. prepared Metis data and performed a Doppler dimming analysis. D.H.B. carried out the EIS observation and provided density and FIP bias results. R.L. performed magnetic field extrapolation and calculated the squashing factor using the PSI-MAS model. D.S. proposed and led the Metis observation in the CH boundary expansion Solar Orbiter Observing Plan. D.B. performed double Gaussian fitting to EIS spectra. L.A., L.M.G., D.M.L., S.L.Y. and A.W.J. aided in the data analysis, discussion of the results and editing of the manuscript text. M.R. is the Metis principal investigator. S.M.G. and A.B. developed Metis data analysis tools and software. F.L. and G.R. were responsible for Metis observation planning for this campaign. All authors reviewed the manuscript.
\end{authorcontribution}

\begin{fundinginformation}
 N.N. is supported by the Science Technology and Facilities Council (STFC) PhD studentship grant No. ST/W507891/1 and University College London (UCL) studentship. The work of D.H.B. was performed under contract to the Naval Research Laboratory and was funded by the NASA Hinode program. R.L. was supported by the NASA Heliophysics Living With a Star Science and Strategic Capabilities programs grant No. 80NSSC20K0192. D.B. is funded under Solar Orbiter EUI Operations grant No. ST/X002012/1 and Hinode Ops Continuation 2022-25 grant No. ST/X002063/1. S.L.Y. would like to thank the Science Technology and Facilities Council for the award of an Ernest Rutherford Fellowship (ST/X003787/1). A.W.J. acknowledges funding from the STFC consolidated grant ST/W001004/1.
\end{fundinginformation}

\begin{dataavailability}
Solar Orbiter data are publicly available at the Solar Orbiter Archive (\url{https://soar.esac.esa.int/soar}). Processed Hinode/EIS data are downloaded from the U.S. Naval Research Laboratory database (\url{https://eis.nrl.navy.mil}). SDO/AIA and SDO/HMI data can be accessed from the Joint Science Operations Center (\url{http://jsoc.stanford.edu}). 
\end{dataavailability}

% \begin{materialsavailability}
% Information about available material ...
% \end{materialsavailability}

% \begin{codeavailability}
% Information about available code ...
% \end{codeavailability}

\begin{ethics}
\begin{conflict}
The authors declare that they have no conflict of interest.
\end{conflict}
\end{ethics}

%%% BIBLIOGRAPHY %%%%%%%%%%%%%%%%%%%%%%%%%%%%%%%%%%%%%%%%%%%%%%%%%%%%%%%%%%%
     % format of references provided by the journal (.bst)
\bibliographystyle{spr-mp-sola}
     % name your Bibtex file containing your references (.bib)
\bibliography{bibliography}  

     % Checking: look if the file containing the ``\bibitem'' exits
     %           so check if the .bbl file exist (bibTeX compilation)
\IfFileExists{\jobname.bbl}{} {\typeout{}
\typeout{****************************************************}
\typeout{****************************************************}
\typeout{** Please run "bibtex \jobname" to obtain} \typeout{**
the bibliography and then re-run LaTeX} \typeout{** twice to fix
the references !}
\typeout{****************************************************}
\typeout{****************************************************}
\typeout{}}

\end{document}